\documentclass[preprint,floatfix,showpacs,byrevtex]{revtex4}
\usepackage{epsfig}
\usepackage{graphics}
\usepackage{amsmath}

\newcommand{\be}{\begin{equation}}
\newcommand{\ee}{\end{equation}}
\newcommand{\bea}{\begin{eqnarray}}
\newcommand{\eea}{\end{eqnarray}}

\newcommand{\slh}{\!\!\!\slash}

\begin{document}

%%%%%%%%%%%%%%%%%%%%%%%%%%%%%%%%%%%%

%\draft
\preprint{\vbox{\hfill \rm ADP-04-01/T580}}
\preprint{\vbox{\hfill \rm UK/04-17}}
%\preprint{ADP-03-106/T544}

\title{Nonperturbative renormalization of composite operators with
overlap fermions}

\author{J.B.~Zhang$^{1,2}$, N.~Mathur$^{3,4}$, S.J.~Dong$^3$, T.~Draper$^3$, I.~Horv\'{a}th$^{3}$, F.X.~Lee$^{5}$,
 D.B.~Leinweber$^2$, K.F.~Liu$^3$, and A.G.~Williams$^2$}

\affiliation{$^1$ Zhejiang Institute of Modern Physics and
Department of Physics, Zhejiang University, Hangzhou 310027, P.R.
China } \affiliation{$^2$
        Special Research Center for the
        Subatomic Structure of Matter and Department of Physics                          \\
        University of Adelaide, Adelaide, SA 5005, Australia }
\affiliation{$^3$ Department of Physics and Astronomy, University
of Kentucky, Lexington, KY 40506} \affiliation{$^4$Jefferson Lab,
12000 Jefferson Avenue, Newport News, VA 23606}
\affiliation{$^5$Center for Nuclear Studies, Department of
Physics, George Washington University,
 Washington, DC 20052}

\vspace{2cm}
%\date{\today}

\begin{abstract}
\vspace{1cm}
We compute non-perturbatively the renormalization constants of composite operators
on a quenched $16^3 \times 28 $ lattice with lattice spacing $a$ = 0.20\,fm
for the overlap fermion by using the regularization independent (RI) scheme.
The quenched gauge configurations were generated with the Iwasaki action.
We test the relations $Z_A = Z_V$ and $ Z_S=Z_P$ and find that they agree well \mbox{(less than 1\%)}
above $\mu$ = 1.6 GeV.
%even for our lattice with a coarse lattice spacing.
We also perform a Renormalization Group (RG) analysis at the
next-to-next-to-leading order and match the renormalization
constants to the $\overline{\rm MS}$ scheme. The wave-function
renormalization $Z_{\psi}$ is determined from the vertex function
of the axial current and $Z_A$ from the chiral Ward identity.
Finally, we examine the finite quark mass behavior for the
renormalization factors of the quark bilinear operators. We find
that the $(pa)^2$ errors of the vertex functions are small and the
quark mass dependence of the renormalization factors to be quite
weak.

\end{abstract}


\maketitle

\section{Introduction}

Lattice QCD is a unique tool to compute the mass
spectrum, leptonic decay constants and hadronic matrix elements of
local operators non-perturbatively from first principles.
Renormalization of lattice operators is an essential ingredient needed
to deduce physical results from numerical simulations.  In this paper
we study the renormalization properties of composite bilinear
operators with the overlap quark action.

In principle, renormalization of quark bilinears can be computed by
lattice perturbation theory.
%However, lattice perturbation theory
%converges slowly and the expansion parameter, the square of the
%lattice coupling evaluated at the lattice scale, $g_0(a)^2$, decreases
%only as an inverse power of $\ln(a)$, hence the higher-order corrections
%may not be small, thus introducing a large uncertainty in the
%calculation of the renormalized matrix elements in some continuum
%scheme.  %This makes systematic improvement of perturbative results very hard.
However, it is generally difficult to go beyond one loop in such
calculations.  To overcome these difficulties, Martinelli ${\it et\, al.}$~\cite{NPM}
have proposed a promising non-perturbative
renormalization procedure.  The procedure allows a full
non-perturbative computation of the matrix elements of composite
operators in the Regularization Independent (RI) scheme
\cite{NPM,eps'/eps}. The matching between the RI scheme and $\overline{\rm MS}$,
which is intrinsically perturbative, is computed using only the well behaved continuum
perturbation theory.

This method has been shown to be quite successful in reproducing
results obtained by other methods, such as chiral Ward Identities
\cite{Ward}.  The method has also been successfully applied to
determine renormalization coefficients for various operators using the
Wilson~\cite{noi_mq,DS99,DS992,becirevic98},
staggered \cite{JLQCD1}, domain-wall~\cite{Tblum1}, chirally
improved~\cite{ggh04}, and overlap fermions~\cite{GHR,zll04}.
The purpose of the current work is to study the
application of this non-perturbative renormalization procedure to the
renormalization of the quark field and
the flavor non-singlet fermion bilinear operators, and also to study their quark mass dependence
for the case of overlap fermions.

Neuberger's overlap fermion~\cite{neub1,neub2} is shown to have correct anomaly
and exact chiral symmetry on
the lattice~\cite{neub1,neub2,luscher} with finite cut-off. As a consequence, many
chiral-symmetry relations~\cite{edwards2,luscher} and the quark
propagator~\cite{liu02} preserve the same structure as in the continuum.
The use of the overlap action entails many theoretical advantages~\cite{neub3}: it has
no additive mass renormalization, there are no $O(a)$ artifacts,
and it has good scaling behavior with small $O(a^2)$ and $O(m^2a^2)$
errors~\cite{kent1,liu02}. In addition, it forbids mixing among operators
of different chirality and, therefore, can be very helpful in computing weak matrix elements.

The outline of this paper is as follows.  In Sec.~\ref{sec:NPRM}, we
review the non-perturbative method (NPM) proposed in Ref.~\cite{NPM}
and introduce the notation used in the remainder of this work.  In
Sec.~\ref{sec:overlap}, we briefly describe the overlap fermion
formalism.  We present the numerical results for the renormalization
constants as well as the Renormalization Group (RG) analysis of
the quark bilinear in Sec.~\ref{sec:numerical} and Sec.~\ref{sec:running}.
In Sec.~\ref{sec:MA}, we examine the finite $m$ behavior of the renormalization
factors of the quark bilinear operators.
We complete our discussion with our conclusions in Sec.~\ref{sec:conclusions}.

\section{Non-perturbative renormalization method}
\label{sec:NPRM}

  In this section, we review the nonperturbative renormalization method of Ref.~\cite{NPM},
which we will use to compute the renormalization constants of quark bilinears in this paper.
The method imposes renormalization conditions non-perturbatively, directly on quark and
gluon Green's functions in Landau gauge.
%Without gauge fixing, all Green functions computed between
%quark and gluon external states are zero.

We start by
considering the definition for the momentum space quark propagator.
Let $S\left(x,0\right)$ be the quark propagator on the gauge-fixed configuration
from a source $0$ to all space-time points $x$. The momentum space propagator is
defined as the discrete Fourier transform over the sink positions
\begin{equation}
S\left(p,0\right) =
\sum_x \exp \left( -i p^{{\rm latt}}\cdot x \right) S \left( x, 0 \right) \, ,
\end{equation}
where $p^{{\rm latt}}$ is the dimensionless lattice momenta.
%\begin{equation}
%p^{\mathrm{latt}}_\mu = \frac{2 \pi}{N_\mu}n_\mu \, ,
%\label{num:mom}
%\end{equation}
%where $\mu$ is one of $x, y, z$ or $t$ and $n_{\mu}$ may in principle
%lie in the range $ 0 \rightarrow N_\mu -1 $ for periodic boundary condition.
%In practice, however, only a subset of this range is used.

   In our case, we use the  periodic boundary condition in spatial directions and the
anti-periodic boundary condition in time direction. We have then the dimensionful momenta
\begin{eqnarray}   \label{eq:pi}
{ p_i = \frac{2\pi}{N_s a} (n_i - N_i/2), ~~~~{\rm and} ~~~~~
p_t = \frac{2\pi}{N_t a} (n_t -1/2 - N_t/2)\, ,}
\end{eqnarray}
for an $N_s^3 \times N_t$ lattice.

We also define the square of the absolute
momentum as the Euclidean inner product
of the momenta defined in Eq.~(\ref{eq:pi})
\begin{equation}
\left( pa\right)^2 = \sum_{\mu} p^{\mathrm{latt}}_\mu \, p^{\mathrm{latt}}_\mu
\, ,
\end{equation}
where we use convention that $p$ is dimensionful and $ p^{\mathrm{latt}}_\mu$ is
dimensionless.

\subsection{ Three Point Function}

Consider the flavor non-singlet fermion bilinear operator
\begin{equation}  \label{bilinear}
O_{\Gamma}(x) = \bar{\psi}(x) \Gamma \psi,
\end{equation}
where $\Gamma$ is the Dirac gamma matrix
\begin{equation}
\Gamma \in \left\{ 1 , \gamma_\mu , \gamma_5, \gamma_\mu
\gamma_5, \sigma_{\mu \nu} \right\} \, ,
\end{equation}
and the corresponding notation will be \{S, V, P, A, T\} respectively. The flavor index
is suppressed. The connected three point function with an operator insertion at position $0$
between the quark fields at $x$ and $y$ is given by
\begin{equation}  \label{three-point}
G_O (x, 0, y) = \langle \psi(x) O_{\Gamma}(0) \bar{\psi}(y)\rangle
=\langle S(x,0) \Gamma S(0,y) \rangle,
\end{equation}
where $S(0,y)$ is the quark propagator from $y$ to $0$. It is the inverse of the
Dirac operator~\footnote{Note that since we use the overlap fermion formalism,
the fermion field $\psi$ in Eqs.~(\ref{bilinear}) and (\ref{three-point}) will
be replaced by $\hat{\psi} = (1 - D/2)\psi$, where $D$ is the massless overlap operator.
As a result, the $S(x,0)$ and $S(0,y)$ are effective quark propagators to be described in
Sec.~\ref{sec:overlap}.}.
Note, $S(x,0)$ here is not translational invariant. Only when averaging over
all gauge configurations, i.e.,
\begin{equation}
\langle S(x,0)\rangle \, ,
\end{equation}
is it translational invariant.

Using $\gamma_5$ hermiticity, the Fourier transform of the three-point function is given by
\begin{eqnarray}
G_O(pa,p^\prime a) &\equiv& \int d^4 x d^4 y e^{-i (p \cdot x - p^\prime \cdot y)} G_O (x, 0, y), \nonumber \\
  & = & \langle \left(\int d^4 x S(x,0) e^{- i p \cdot x}\right) \Gamma
                \left(\int d^4 y S(0,y) e^{ i p^\prime \cdot y} \right) \rangle, \nonumber \\
  & = & \langle S(p,0) \Gamma \gamma_5 \left(\int d^4 y S^{\dagger}(y,0)
         e^{i p^\prime \cdot y}\right)\gamma_5 \rangle,
\end{eqnarray}
where the $\dagger$ refers only to the color and spin indices.
This can be written as
\begin{equation}
G_O(pa,p^\prime a)   =  \langle S(p,0) \Gamma  \left(\gamma_5 S^{\dagger}(p^\prime,0)\gamma_5\right)\rangle,
\end{equation}
From this, one can define the vertex function as the amputated
three-point function
\begin{equation}  \label{lambda_pq}
\Lambda_O (pa, p^\prime a) = S(pa)^{-1} G_O(pa,p^\prime a) S(p^\prime a)^{-1},
\end{equation}
\label{lambdao}
where
\begin{equation}
S(pa) = \langle S(p,0) \rangle,
\end{equation}
which is translational invariant and is a $12 \times 12$ matrix in color-spin space.

   Finally, a projected vertex function is defined
\begin{equation}
\Gamma_O(pa) = \frac{1}{\mathrm{Tr}(\hat{P}_O^2)} \mathrm{Tr} \left(\Lambda_O (pa, pa) \hat{P}_O\right),
\label{eqvertex}
\end{equation}
where $\hat{P}_O = \Gamma$ is the corresponding projection operator.

\subsection{RI-MOM Renormalization Condition}   \label{RI/MOM}

The renormalized operator $O(\mu)$ is related to the bare operator
\begin{equation}
O(\mu) = Z_O(\mu a, g(a)) O(a) \, ,
\end{equation}
and the renormalization condition is imposed on the three-point vertex function $\Gamma_O(pa)$
at a scale $p^2 = \mu^2$ as
\begin{equation}    \label{eq:ren_con}
\Gamma_{O, {\rm ren}}(pa)|_{p^2 = \mu^2} = \frac{Z_O(\mu a, g(a))}{Z_{\psi}(\mu a, g(a))}
\Gamma_O(pa)|_{p^2 = \mu^2} = 1
\end{equation}
to make it agree with the tree-level value of unity~\cite{NPM}. Here $Z_{\psi}$ is the
field or wave-function renormalization
\begin{equation}
\psi_{\rm ren} = Z_{\psi}^{1/2} \psi.
\end{equation}

   In order to alleviate the non-perturbative effects from the spontaneous chiral symmetry breaking,
high virtuality with $\mu \gg \Lambda_{QCD}$ is required. On the
other hand, to avoid the discretization errors, one would need
$\mu \ll 1/a$. So, for the RI-MOM procedure to be a valid and
practical renormalization scheme, there should exists a window in
the renormalization scale $\mu$, i.e. $\Lambda_{QCD} \ll \mu \ll
1/a$. In this work, the renormalization constants are extracted
from $(pa)^2 > 1$ which, in principle, should have large
discretization errors. But, as we will see later, we do not see
large discretization errors. This is also the case observed in
previous studies~\cite{NPM,noi_mq,Tblum1,zll04}.

In practice, one usually matches the results to the perturbative
scheme, e.g. $\overline{\rm MS}$ scheme, in order to compare with
experimental quantities. We will discuss perturbative matching in
Sec.~\ref{sec:numerical}. In general, the vertex function
$\Gamma_O(pa)$ may have intrinsically non-perturbative
contributions, e.g.~from the Goldstone boson propagator, which are
not included in perturbative calculations. To this end, we either
go to large enough momentum such that the non-perturbative effects
are suppressed or somehow remove them from the data.

   There are several ways to obtain the renormalization constant $Z_O$ for the operator $O$. For the
$Z_{\psi}$ in Eq. (\ref{eq:ren_con}), one could use a known ratio
involving $Z_{\psi}$ to have it eliminated in Eq.
(\ref{eq:ren_con}). The first way is to extract $Z_{\psi}$ from
the vertex function of the conserved vector or axial-vector
current. For example, if one uses the conserved vector current,
then $Z_{V_C} = 1$. From the renormalization condition
\begin{equation}
\frac{Z_{V_C}}{Z_{\psi}} \Gamma_{V_C}(pa)|_{p^2 = \mu^2} = 1,
\end{equation}
one then obtains
\begin{equation}
Z_{\psi} = \frac{1}{48} {\mathrm{Tr}}\left(\Lambda_{V_{C\mu}}(pa) \gamma_{\mu}\right)|_{p^2 = \mu^2}.
\end{equation}

   One can also extract $Z_{\psi}$ directly from the quark propagator.
From Ward Identity (WI), it follows~\cite{NPM}
\begin{equation}
Z_{\psi}= \frac{-i}{12} {\mathrm{Tr}} \left[\frac{\partial S(pa)^{-1}}{\partial \slh{p}}\right]
_{p^2=\mu^2}\; .
\label{eq:Z_q_WI}
\end{equation}
To avoid derivatives with respect to a discrete variable, it is suggested~\cite{NPM} to use

\begin{equation}  \label{wf_ren}
Z'_{\psi}=\left.\frac{-i}{12}\frac{{\mathrm{Tr}} \sum_{\mu=1}^4
\gamma_\mu (p_\mu a)S(pa)^{-1}}
{4\sum_{\mu=1}^4 (p_\mu a)^2}\right|_{p^2=\mu^2}\; ,
%\label{eq:Z_q'_WI}
\end{equation}
which, in the Landau gauge, differs from $Z_\psi$ by a finite term of order
$\alpha_s^2$~\cite{Franco:1998bm}. The matching coefficient can be computed using continuum
perturbation theory, and up to order $\alpha_s^2$~\cite{Franco:1998bm}
\begin{equation}
\frac{Z_\psi^\prime}{Z_\psi}=
1-\dfrac{\alpha_s^2}{\left( 4\pi \right) ^2}\Delta _\psi^{(2)}+\ldots
\label{eq:deltaq1}
\end{equation}
In the Landau gauge
\begin{equation}
\Delta^{(2)}_\psi =
 {{\left( N_c^2 - 1 \right) }\over {16\,N_c^2}} \,
 \left(3 + 22 N_c^2 - 4 N_c n_f \right),
\label{eq:deltaq2}
\end{equation}
where $N_c$ is the number of colors and $n_f$ the number of dynamical quarks.
However, as pointed out in Ref.~\cite{Tblum1}, due to the ambiguity as to how the discrete lattice
momentum $p$ is defined, this method will introduce roughly 10\% -- 20\% uncertainty in determining
$Z_{\psi}$.

   The third way is to first calculate the renormalization constant $Z_A$ from
the axial Ward identity~\cite{kent1,GHR,kent2} with a local current, and then use the
renormalization condition for the axial current
\begin{equation}   \label{Z_A}
\frac{Z_A}{Z_{\psi}} \Gamma_{A}(pa)|_{p^2 = \mu^2} = 1,
\end{equation}
to eliminate the unknown $Z_{\psi}$ from Eq. (\ref{eq:ren_con}). Combining Eqs. (\ref{eq:ren_con}),
(\ref{Z_A}) and the $Z_A$ from the Ward identity, one obtains
\begin{equation}  \label{3rd}
Z_O = Z_A \frac{\Gamma_{A}(pa)|_{p^2 = \mu^2}}{\Gamma_{O}(pa)|_{p^2 = \mu^2}}.
\end{equation}
This way, other renormalization constants, such as $Z_S$, $Z_P$, $Z_V$, and $Z_T$ can be obtained
and the identity relations $Z_S = Z_P$ and $Z_V = Z_A$ due to chiral symmetry can be checked.
Note here, the local axial-vector current is finite. Thus, $Z_A$ is independent of scale, but depends
on the lattice spacing $a$.
%We expect it (and the  $Z_V$ of the local vector current) to go to unity at the
%continuum limit.

In this work, we shall adopt the third approach as mentioned above.
It is known that $Z_A$ as determined from the Ward identity has a small
statistical error at the level of 0.2\%~\cite{kent2} and will not contribute
much to the overall error. When we study the quark mass dependence,
we shall use the wave-function renormalization from the quark propagator to obtain its
dependence for the renormalization factors of the quark bilinear operators.

\section{overlap fermion}
\label{sec:overlap}

%### I have rewritten the formalism for this section with the correct
%and preferred factors and not to follow the cumbersome way that is in
%our notes and the computer codes. Let's not confuse the readers with
%our idiosyncrasy.
%%%%%%%%%%%%%%%%%%%%%%%%%%%%%%%%%%%%%%%%%%%%%%%%

The massless overlap-Dirac operator in
lattice units is~\cite{neub2}
\begin{equation}
   D(0) = \left[1 + \gamma_5 \epsilon(H)\right] \, ,
\label{D0_definition}
\end{equation}
where $\epsilon(H)$ is the matrix sign function of an Hermitian operator $H$.
$\epsilon(H)$ depends on the background gauge field and has eigenvalues $\pm 1$.
Any such $D$ is easily seen to satisfy the Ginsparg--Wilson
relation~\cite{GW}
\begin{eqnarray}
\{\gamma_{5},D\}=D\gamma_{5}D \, .
\label{ginspargrel}
\end{eqnarray}
For the topological sector with no zero modes, it follows easily that
$\{\gamma_5,D^{-1}(0)\}=\gamma_5$ and by defining
$\tilde D^{-1}(0)\equiv [D^{-1}(0)-1/2]$ we see that it anticommutes with $\gamma_5$
\begin{equation}
\{\gamma_5,\tilde D^{-1}(0)\}=0.
\end{equation}

The standard choice of $\epsilon(H)(x,y)$ is $\epsilon(H)\equiv
H_W/|H_W| = H_W/(H_W^\dagger H_W)^{1/2}$, where $H_W(x,y)=\gamma_5
D_W(x,y)$ is the Hermitian Wilson-Dirac operator. $D_W$ is the
usual Wilson-Dirac operator on the lattice.  However, in the
overlap formalism the Wilson mass parameter $\rho$ needs to be
negative in order to generate zero modes.

In the present work, we use
the standard Wilson-Dirac operator, which can be written as
\begin{equation}
D_{W}(x,y)
= \left[\delta_{x,y}
-\kappa\sum_\mu
\left\{(r-\gamma_\mu)
{U_\mu(x)}\delta_{y,x+\hat\mu}+(r+\gamma_\mu)
{U^\dagger_\mu(x-a\hat{\mu})}\delta_{y,x-\hat\mu}
\right\}
\right] \, .
\label{diracop}
\end{equation}
The negative Wilson mass $-\rho $ is then related to $\kappa$ by
\begin{equation}
\kappa\equiv\frac{1}{2(-\rho) +~8 } \, .
\label{kappa_defn}
\end{equation}
$\rho$ is chosen such that $\kappa > \kappa_c$ and $\rho < 2 r$,
and hence there is no species doubling and there can be zero modes
for the  massless quark. We take $r=1$ and $\rho = 1.368$ (which
corresponds to $\kappa$ = 0.19) in our numerical simulation.

It is shown that the flavor non-singlet scalar,
pseudo-scalar~\cite{hhh02}, vector, and axial~\cite{ky99,hhh02}
bilinears in the form $\bar{\psi}KT (1 - \frac{1}{2}D)\psi$ ($K$ is
the kernel which includes $\gamma$ matrices and $T$ is the flavor $SU(N_f)$ matrix) transform
covariantly under the global chiral transformation $\delta \psi = T\gamma_5(1 - D/2)\psi$
as in the continuum. The $1 - \frac{1}{2}D$ factor is also understood as
the lattice regulator which projects out the unphysical real
eigenmodes of $D$ at $\lambda = 2$.  For the massive case, the fermion action
is defined as $\bar{\psi}\rho D \psi + m a \bar{\psi}(1 - \frac{1}{2}D) \psi$
so that the tree-level wave-function renormalization of the quark propagator is unity.
In this case, the Dirac operator can be written as
\begin{equation}  \label{D_m}
D(m)=  \rho D + m a (1 - \frac{1}{2} D) = \rho+\frac{m}{2} + (\rho-\frac{m}{2})\gamma_5\epsilon(H).
\end{equation}

With the $\psi$ field in the operators and the interpolation fields for hadrons
replaced by the lattice regulated field $\hat{\psi} = (1 -
\frac{1}{2}D)\psi$, the regulator factor will be associated with the
quark propagator in the combination $(1 - \frac{1}{2}D) D(m)^{-1}$ in
Green's functions, leading to an effective quark propagator~\cite{liu02}
\begin{equation}  \label{propagator}
S(x,y) = (1 - \frac{1}{2}D) D(m)^{-1}
= ( D_c + m )^{-1}.
\end{equation}
where the operator $D_c =  \rho D/(1 - \frac{1}{2}D)$ is chirally
symmetric in the continuum sense, i.e. $\{\gamma_5, D_c\} =
0$~\cite{cz98,neu98}; but, unlike $D$, it is non-local. Thus, the
effective quark propagator in Eq.~(\ref{propagator}) turns out to
have the same form as in the continuum, i.e. a chirally symmetric
$D_c$ plus a mass term in the inverse
propagator~\cite{liu02,chiu99,cgh99,overlgp}. By studying the
dispersion relation of the pseudoscalar and vector mesons, it is
learned~\cite{liu02} that the $(ma)^2$ errors are much smaller
than those of the Wilson fermion, making it a viable option for
studying the heavy-light systems. Furthermore, it affords a
non-perturbative renormalization of the heavy-light decay constant
via the chiral Ward identity and the unequal mass
Gell-Mann-Oakes-Renner relation~\cite{liu02}. The preliminary
study of the charmonium spectrum with the quenched overlap fermion
seems encouraging as far as the hyperfine splitting and the S-wave
to P-wave charmonium splittings are concerned~\cite{tac05}.

\section{NUMERICAL RESULTS}
\label{sec:numerical}

 In this paper we work on a $16^3\times{28}$ lattice with lattice spacing, $a$=0.20\,fm,
as determined from the pion decay constant $f_\pi$~\cite{kent3}.
The gauge configurations are created by the Iwasaki
gauge action through the pseudo-heat-bath algorithm. A total
of 80 configurations are used.
The lattice parameters are summarized in Table~\ref{simultab}.

\begin{table}[ht]
\caption{\label{simultab}Lattice parameters.}
\begin{ruledtabular}
\begin{tabular}{cccccccc}
Action &Volume &$N_{\rm{Therm}}$ & $N_{\rm{Samp}}$ &$\beta$ &$a$ (fm) & Physical Volume (fm$^4$)\\
\hline
Iwasaki       & $16^3\times{28}$ & 10000 & 5000 & 2.264 & 0.200 & $3.2^3\times{5.60}$ \\
%\hline
\end{tabular}
\end{ruledtabular}
\end{table}

Recently it has been speculated~\cite{gss05} that the overlap
operator with coarse lattice spacing of 0.2 fm, such as in
Ref.~\cite{kent3} and this study, might have a range as large as 4
lattice units. Thus, the calculations might be afflicted by
unphysical degrees of freedom as light as 0.25 GeV. It has been
shown~\cite{dmz05} by direct calculations at lattice spacings of
0.2 fm, 0.17 fm, and 0.13 fm that this is not the case. The range
of the overlap operator as defined in Ref.~\cite{hjl99} is about 1
lattice unit (in Euclidean distance or 2 units of `taxi driver'
distance) for each of the above lattice spacings. Therefore, the
range of the overlap Dirac operator scales to zero in physical
units in the continuum limit and our $16^3 \times 28$ lattice,
with spacing of 0.2 fm, used in this study is in the scaling
range.

The gauge field configurations are gauge fixed to the Landau gauge
using a Conjugate Gradient Fourier Acceleration~\cite{cm}
algorithm with an accuracy of
$\theta\equiv\sum\left|\partial_{\mu}A_{\mu}(x)\right|^{2}<10^{-12}$.
We use an improved gauge-fixing scheme~\cite{bowman2} to minimize
gauge-fixing discretization errors. Since gauge-fixing is involved
for the external quark state, there are concerns about the effects
of Gribov copies on the numerical results of the renormalization
procedure. A study of two different realization of the Landau
gauge and a covariant gauge shows that the renormalization
constants from these gauge-fixings differ by less than the
statistical errors of about 1 -- 1.5\% level~\cite{gpt02}. We
shall thus assume that that the potential uncertainty due to
Gribov copies is at this 1 -- 1.5\% level which is comparable to
our statistical errors.

Our numerical calculation begins with an evaluation of the inverse
of $D(m)$ which is defined in Eq.~(\ref{D_m}). We use a 14th-order
Zolotarev approximation~\cite{zolo} to the matrix sign function
$\epsilon(H_W)$. In the selected window of $x \in [0.031,2.5] $ of
$\epsilon(x)$, the approximation  is better than $3.3 \times
10^{-10}$~\cite{kent3}. We then calculate Eq.~(\ref{propagator})
for each configuration by using multi-mass Conjugate Gradient
method for both the inner and outer loops. The detailed numerical
description is given in Ref.~\cite{kent3}.
In the calculations, $\kappa=0.19$  was used, which corresponds to $\rho = 1.368$.
We calculate 15 quark masses by using a shifted version of
the Conjugate Gradient solver~\cite{edwards2,kent1}.
The bare quark masses $ma$ are chosen to be
$ma$ = $0.02100$, $0.03033$, $0.04433$, $0.06417$, $0.07583$, $0.08983$, $0.10850$, $0.12950$,
$0.15633$, $0.18783$, $0.22633$, $0.26833$, $0.32200$, $0.40000$, and  $0.60000$.
With the scale determined by $f_{\pi}$, they correspond to pion masses 212(7), 247(6), 290(6),
342(6), 370(7), 400(7), 438(7), 478(8), 524(8), 575(9), 633(10), 692(11), 764(12), 862(13),
1092(17)~MeV respectively~\cite{kent3}.

%With the renormalization constants to be calculated later included, the
%$\overline{\rm MS}$ masses at $\mu = 2$ GeV will be about half of these bare
%masses.

In the following, we give the steps for the numerical calculation:

\begin{itemize}
\item
 After we calculate the quark propagators in coordinate space for each
 configuration, we use the Landau gauge transformation matrix
 to rotate the quark propagators to the Landau gauge. The discrete
 Fourier transform is then used to calculate the quark propagators in
 momentum space.

\item
Next, we calculate the five projected vertex functions $ \Gamma_O(pa) $
defined in Eq.~(\ref{eqvertex}), where we have used the effective quark
propagator in Eq.~(\ref{propagator}) for the calculation.  By
definition, they are the ratios of renormalization constants at
the chiral limit (i.e. $Z_{\psi}(\mu a, g(a))/Z_O(\mu a, g(a))$
from Eq.~(\ref{eq:ren_con})) in the RI scheme at scale $\mu^2=p^2$.
They are in general dependent on $(pa)^2$ which comes from two
sources. One is from the running of the renormalization constants in the RI
scheme; the other is from the possible $(pa)^2$ error.

\item
 We decouple the two above mentioned scale dependencies of the
 calculated $ \Gamma_O(pa) $ by first dividing out its perturbative
 running in the RI scheme. Ideally, this should take care of the scale
 dependence, since we have taken the scale to infinity. However, due
 to the $(pa)^2$ error on the lattice, there can still be some $(pa)^2$
 dependence in the $\Gamma_O(pa)$ after undoing the perturbative
 running. Following Ref.~\cite{Tblum1}, we shall attribute the remaining
 scale dependence to the $(pa)^2$ error and will use the simple linear
 fit to remove it. This will be discussed in Section~\ref{sec:running}. For the
 scalar and pseudo-scalar vertex functions, there is an additional
 complication due to the presence of quark mass
 poles~\cite{Tblum1}. We shall remove them first and then extrapolate to the
 chiral limit. Finally, we can check the expected relations $Z_A = Z_V$
 and $Z_S = Z_P$ .

\item
In order to compare results with experiments, one frequently quotes the
scale-dependent results in the $\overline{\rm MS}$ scheme at certain scale. So the
final step is to perturbatively match the results from the RI scheme to the
$\overline{\rm MS}$ scheme at $\mu = 2\,{\rm GeV}$ for $Z_S, Z_P$ $Z_T$, and $Z_{\psi}$.

\end{itemize}

\subsection{Axial and vector currents}

\begin{figure}[tp]
\centering{\epsfig{angle=90,figure=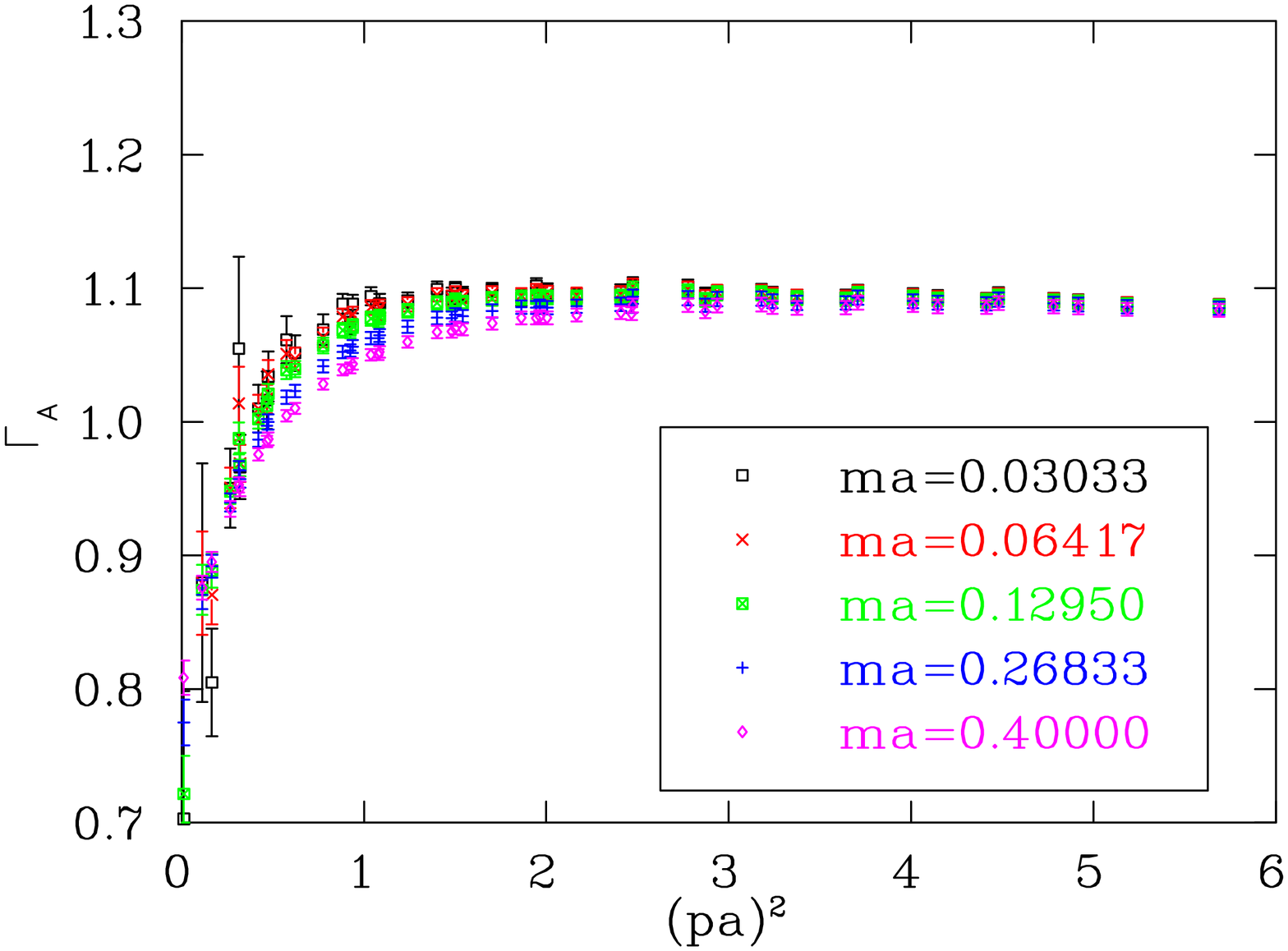,height=8cm} }
\centering{\epsfig{angle=90,figure=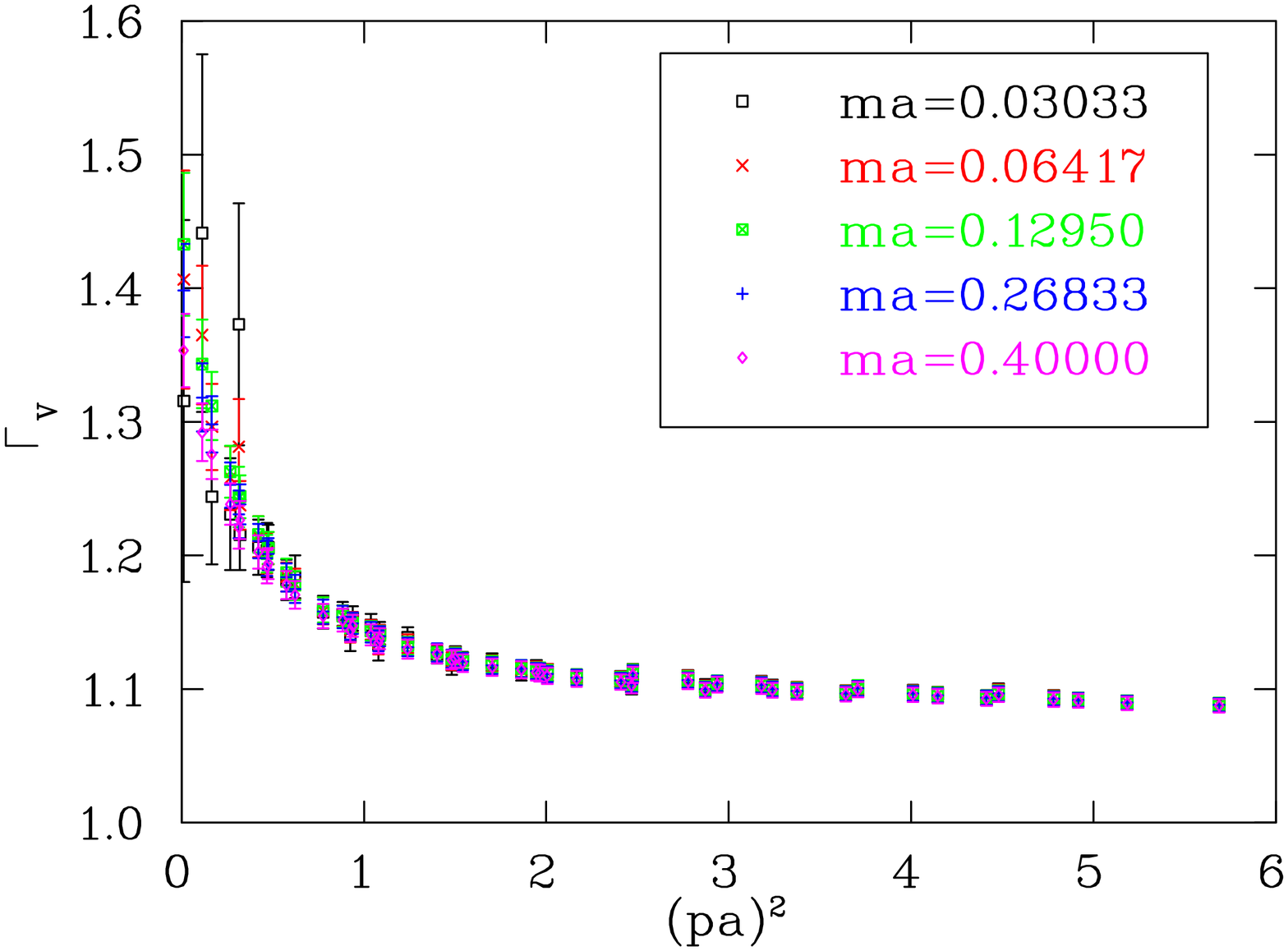,height=8cm} }
\parbox{130mm}{
\caption{The projected vertex function $\Gamma$ defined in Eq.~(\ref{eqvertex})
for the vector and axial-vector currents with different bare quark masses.
% they are the ratios of  renormalization constants
%$Z_{\psi}/Z_A$ and $Z_{\psi}/Z_V$ respectively.
}
\label{figav}}
\end{figure}

\begin{figure}[tp]
\centering{\epsfig{angle=90,figure=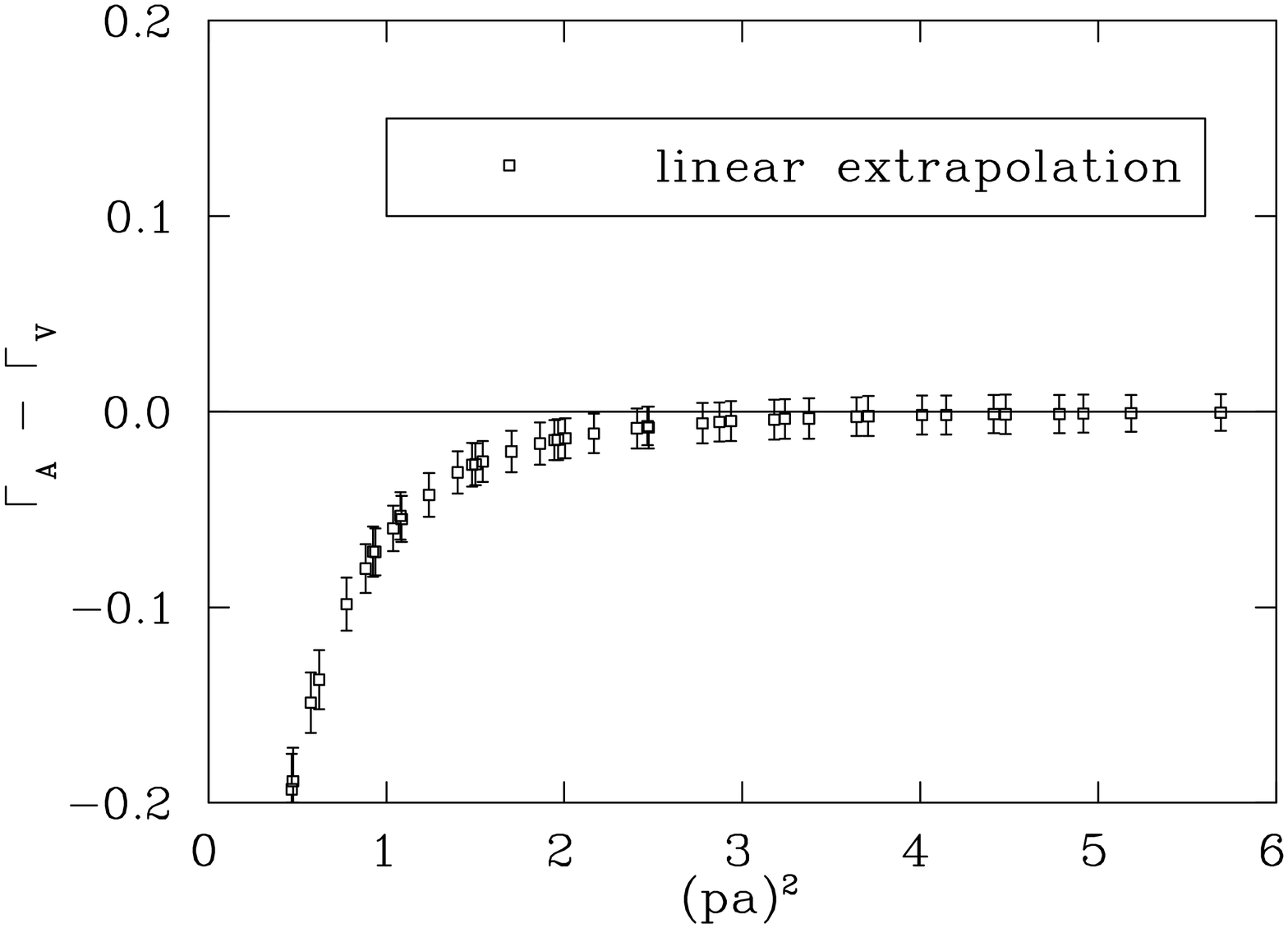,height=8cm} }
\centering{\epsfig{angle=90,figure=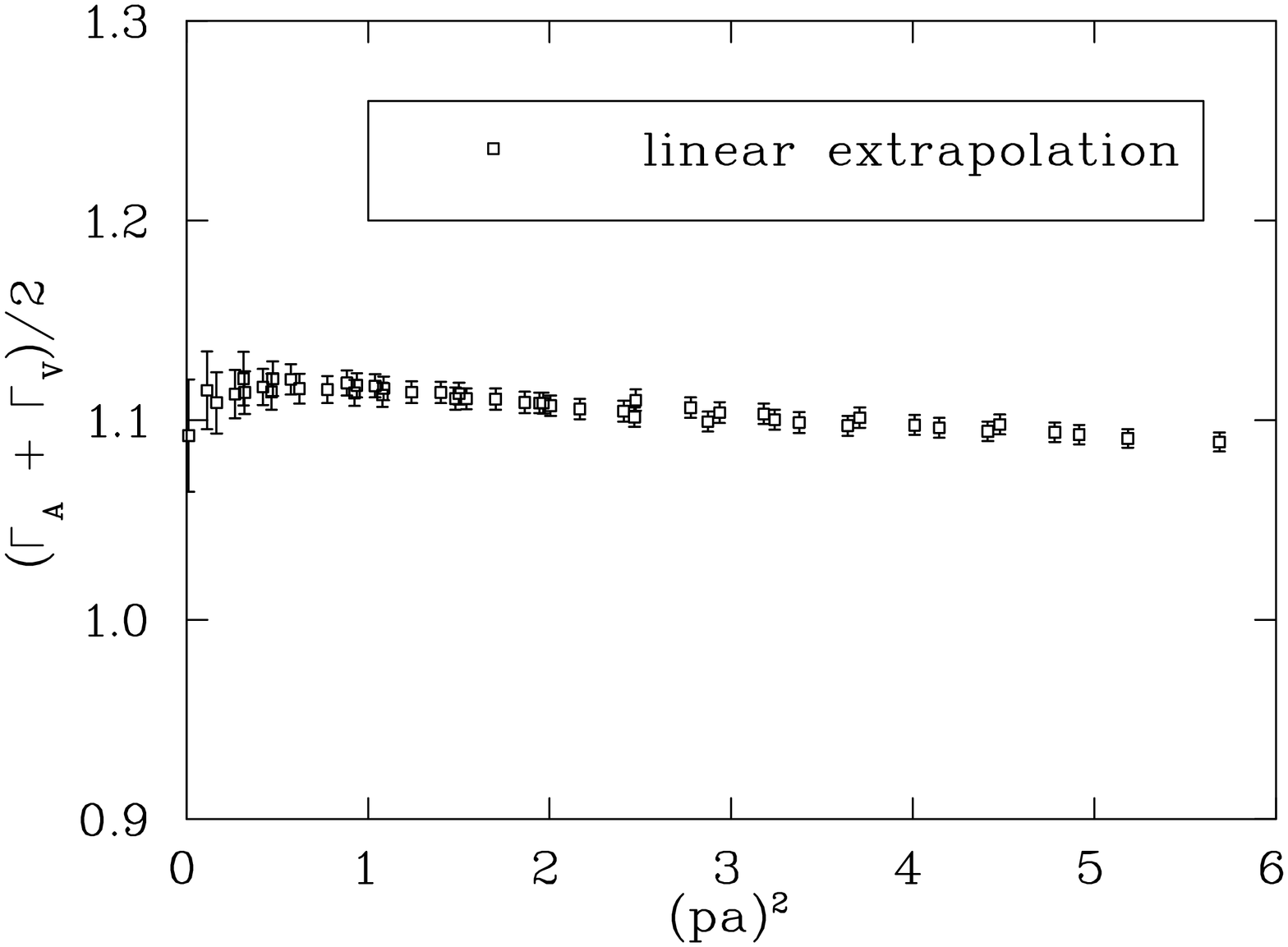,height=8cm} }
\parbox{130mm}{
\caption{The upper panel is ($\Gamma_A-\Gamma_V$)
versus $(pa)^2$. We see that $Z_A$ = $Z_V$ is valid above moderate $(pa)^2 \sim 2.5 $.
%### Question: by linear, I suppose you mean  $(ma)^2$? No, is (ma).
%&&& Why extrapolate with ma? There is no ma error. Should extrapolate
%&&& with $m^2a^2$ and $ma^2$. There is no difference between $ma^2$ and $ma$
%&&& since $a$ is fixed. But I expect the $m^2a^2$ term to be dominating.
Here we linear extrapolate to the chiral limit ($m=0$)
with respect to $ma$.
The lower panel is $\frac{1}{2}$ ($\Gamma_A+\Gamma_V$)
versus $(pa)^2$.}
\label{Checkva}}
\end{figure}

Let us consider first the vector and axial vector currents. Since each obeys a
Ward Identity~\cite{Bochicchio}, their renormalization constants
are finite. In Fig.~\ref{figav}, we show the vertex functions $\Gamma_A$ and $\Gamma_V$
from Eq.~(\ref{eqvertex}) for different bare quark masses as a function of the
lattice momentum $(pa)^2$. We find that they are weakly dependent on the mass,
and almost scale independent after $ (pa)^2~ \ge ~2.0 $, which corresponds to
$p \ge $ 1.4 GeV.

 In the RI scheme, Eq.~(\ref{eq:ren_con}) implies that in the chiral limit
\begin{equation}
\lim_{m \rightarrow 0}\Gamma_{A/V} (pa)|_{p^2 = \mu^2} = Z_{\psi}(\mu a, g(a))/Z_{A/V},
\end{equation}
and one expects that $Z_A = Z_V$ for the overlap fermion, but this is true only for
large momenta $p$.
%it is not guaranteed that $\Gamma_A$ equals $\Gamma_V$ at all scale.
At low momenta, $\Gamma_A$ and $\Gamma_V$ may differ due to
the effects of spontaneous chiral symmetry breaking~\cite{Tblum1}.

Following Ref.~\cite{Tblum1}, we show in Fig.~\ref{Checkva} the
quantities $\Gamma_A - \Gamma_V $ and $\frac{1}{2}(\Gamma_A +
\Gamma_V) $, after linearly extrapolating with respect to $ma$ to
the chiral limit. We can observe from the upper panel of the
figure that there is no effect of spontaneous chiral symmetry
breaking at moderate and high momenta, where $\Gamma_A- \Gamma_V$
tends to zero. In fact, the percentage error between $\Gamma_A$
and $\Gamma_V$ is less than $1\%$  at $(ap)^2=2.5$ (top panel) and
is comparable to the statistical error. This deviation decreases
further for higher momenta. On the other hand, the effects of
spontaneous chiral symmetry breaking are clearly visible at low
momenta where $\Gamma_A$ and $\Gamma_V$ differ. In the lower panel
of Fig.~\ref{Checkva}, we plot $\frac{1}{2}(\Gamma_A + \Gamma_V) $
against $(pa)^2$.
%We shall follow Ref.~\cite{Tblum1} and use
%the sum  $\frac{1}{2}(\Gamma_A + \Gamma_V) $ for the extraction of
%both $Z_A/Z_\psi$ and $Z_V/Z_\psi$ to increase statistics.
Since $Z_V$ and $Z_A$ are scale independent, the slight $(pa)^2$ dependence that one observes
in $\frac{1}{2}(\Gamma_A + \Gamma_V)$ in Fig.~\ref{Checkva} for $(pa)^2 > 2.5$ reflects
the scale dependence of $Z_{\psi}$ and the lattice $(pa)^2$ error.

\subsection{Pseudo-scalar and scalar densities}

The pseudo-scalar and scalar densities differ from the axial and
vector currents in the sense that their renormalization constants
are not scale independent. The scalar and pseudo-scalar densities
with the form $\bar{\psi}(1 - D/2)\psi$ and $\bar{\psi}\gamma_5(1
- D/2)\psi$ transform under the lattice chiral transformation as
in the continuum~\cite{luscher,has98,vicari}. From the Ward
identities, one obtains the relations \bea
Z_S &=& Z_P \\
Z_m &=& \frac{1}{Z_S}.
\eea
Thus, the quantities $Z_S/Z_P$, $Z_SZ_m$ and $Z_m Z_P$ are expected
to be scale independent.

For the case of the pseudo-scalar and scalar renormalization, there is a complication
due to the presence of the quark condensate in the inverse quark propagator. Using the axial
Ward identity from the quark propagator, one has~\cite{Tblum1}

\begin{equation}   \label{Gamma_P}
m \Gamma_P (p,p) = \frac{1}{12} Tr (S^{-1}(p)).
\end{equation}

It is known that due to the spontaneous chiral symmetry breaking, the trace of the inverse
quark propagator picks up a contribution from the quark condensate~\cite{lpp74}.
At large $p^2$~\cite{lpp74}, it is given by
\begin{equation}
 \frac{1}{12} Tr (S^{-1}(p)) = m - \langle \bar{q}q\rangle \frac{4 \pi \alpha_s}{3 p^2}
 + O (1/p^4).
\end{equation}
from first order perturbation~\cite{lpp74}. This implies that the renormalized
$ Tr (S_{ren}^{-1}(p))$ should be
\begin{equation}  \label{tr_S}
\frac{1}{12} Tr (S_{ren}^{-1}(p)) = m_{ren} - C_1 \frac{\langle \bar{q} q \rangle}{p^2} + O(1/p^4) \, .
\end{equation}
In the study of lattice artifacts of the Wilson fermion~\cite{bgl00}, it is shown
that there are three terms which mix at $O(a)$ to give an improved and renormalized
quark field,
\begin{equation}  \label{ren_q}
q_{ren} = Z_{\psi}^{1/2} (1 + b_q ma) \{1 + a c_q'(\not\!\!{D} + m_{ren}) + a c_{NGI} \not\!\partial \} q_0,
\end{equation}
where $\not\!\partial$ may appear due to gauge fixing. It is found~\cite{bgl00} in the study of
the order $O(a)$ error of Wilson fermion, that $c_q'$ is large.
Combining Eqs.~(\ref{tr_S}) and (\ref{ren_q}), one has~\cite{Tblum1}
\begin{equation}  \label{tr_S_lat}
\frac{1}{12} Tr (S_{latt}^{-1}(pa)) = ... + Z_m Z_{\psi} (ma +
m_{res}a) - C_1 Z_{\psi} \frac{a^3 \langle
\bar{q}q\rangle}{(pa)^2} + 2(c_{NGI} - c_q')(pa)^2 + O(1/p^4),
\end{equation}
where terms of $O(mc_{NGI})$ are neglected. Thus, in this case,
$\frac{1}{12} Tr (S_{latt}^{-1}(pa))$ diverges for large $pa$.

    On the other hand, it is learned from the domain wall fermion~\cite{Tblum1} study on a
$16^3 \times 32\times 16$ lattice with Wilson gauge action at $\beta=6.0$, the explicit chiral
symmetry breaking effect from the $c_{NGI} - c_q'$ term is negligible for
moderately large values of $(pa)^2$ and that the residual mass is small. Since
the explicit symmetry breaking is controlled to a level $< 10^{-9}$ with the Zoloterav approximation
of the sign function~\cite{kent3} for the overlap fermion,
we expect the $c_{NGI} - c_q'$ term to be negligibly small. It was already shown
that the residual quark mass due to the numerical approximation of the overlap operator is
negligible~\cite{kent1}. Here, we shall verify the expectation that the $c_{NGI} - c_q'$ term
is indeed negligible.

%### please add the corresponding Fig. 1,2, and 3 in Ref. (Tblum1) in the following.

\begin{figure}[tp]
%\centering{\epsfig{angle=90,figure=./graphs/Bp.combM.psq.nm6.ps,height=8cm} }
\centering{\epsfig{angle=90,figure=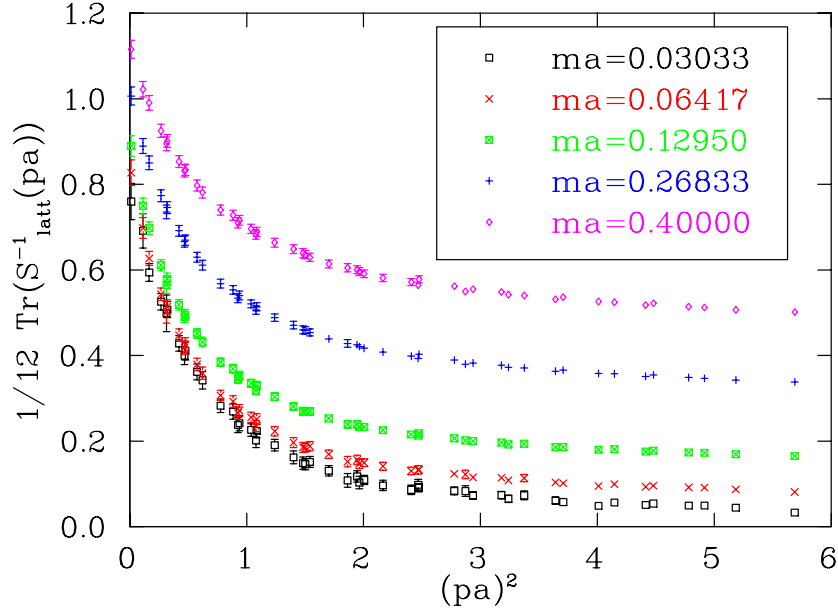,height=8cm} }
%\parbox{130mm}{
\caption{A plot of $\frac{1}{12} Tr (S_{latt}^{-1}(pa))$ versus
$(pa)^2$ for different bare quark mass $ma$ on $16^3 \times 28 $
lattice, showing that $\frac{1}{12} Tr (S_{\mathrm
latt}^{-1}(pa))$ approaches a constant value at large $ (pa)^2$. }
\label{tr_S^{-1}}
\end{figure}

\begin{figure}[tp]
\centering{\epsfig{angle=90,figure=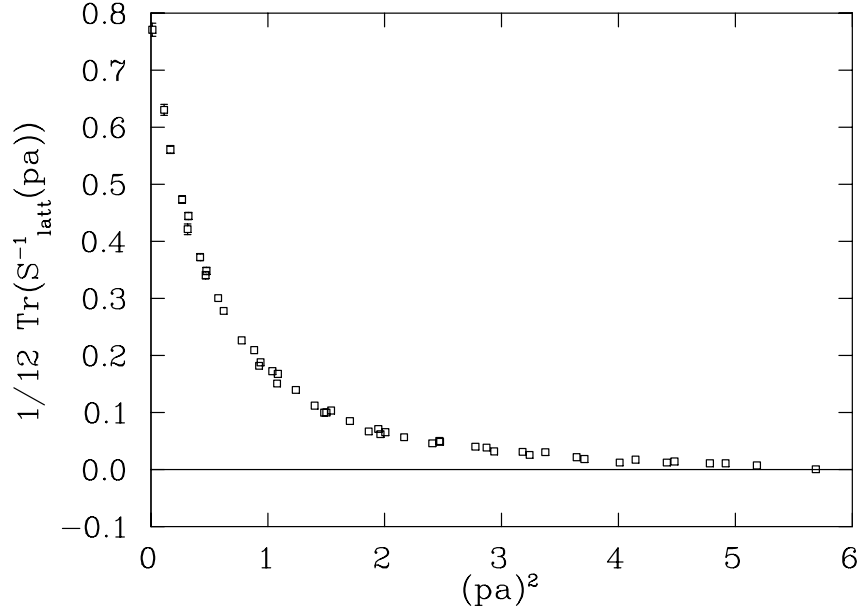,height=8cm} }
%\parbox{130mm}{
\caption{ The value of $\frac{1}{12} Tr (S_{latt}^{-1}(pa))$
extrapolated to $m=0$ vs $ (pa)^2 $ by a
%&&& To be consistent, should use m^2 a^2 or both ma^2 and m^2a^2 to extrapolate!
simple linear extrapolation. At large $(pa)^2$, the extrapolated value is zero within errors. }
\label{tr_S_m0}
\end{figure}

\begin{figure}[tp]
\centering{\epsfig{angle=90,figure=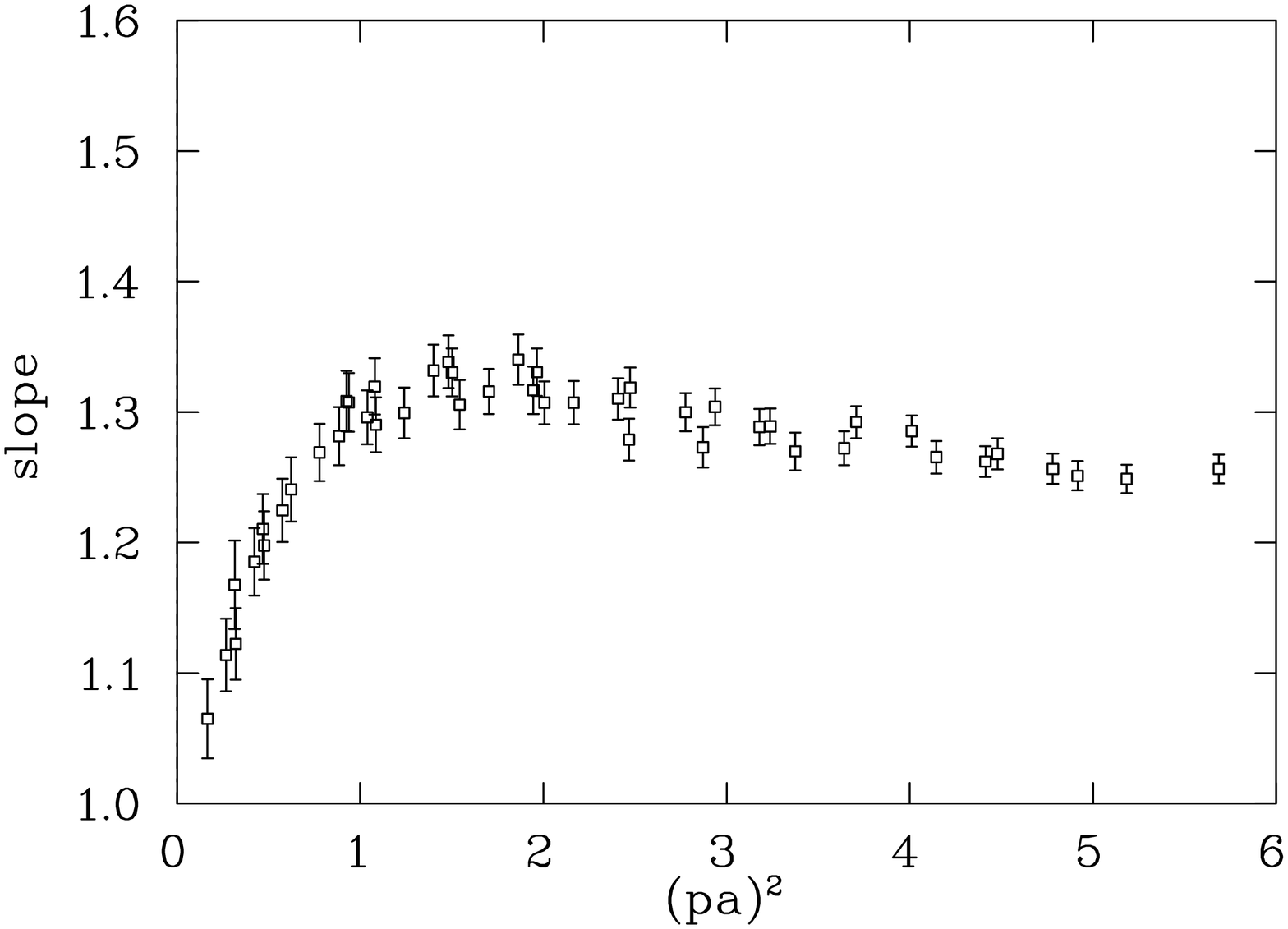,height=8cm} }
%\parbox{130mm}{
\caption{A plot of the slope of $\frac{1}{12} Tr
(S_{latt}^{-1}(pa))$ with respect to the quark mass as a function
of $(pa)^2$. It is expected to be $Z_m Z_{\psi}$ at large $(pa)^2$
from Eq.~(\ref{tr_S_lat}).} \label{ZmZq}
\end{figure}

%Figs. \label{tr_S^{-1} here.
%Fig. \label{tr_S_m0}
%Fig. \label{ZmZq}

   As shown in Fig.~\ref{tr_S^{-1}}, $\frac{1}{12} Tr (S_{latt}^{-1}(pa))$ for several quark
masses tend to constant values after $(pa)^2 \ge 4 $ . This
indicates that there is no discernible contamination due to the
explicit chiral symmetry term $2(c_{NGI} - c_q')(pa)^2$ which
grows as $(pa)^2$. Fig.~\ref{tr_S_m0} shows the same at the chiral
limit which is obtained from linear extrapolation in $ma$. In this
case, $\frac{1}{12} Tr (S_{latt}^{-1}(pa))$ tends to zero at large
$(pa)^2$ as expected from Eq.~(\ref{tr_S_lat}) with no residual
mass. Plotted in Fig.~\ref{ZmZq} is the slope of $\frac{1}{12} Tr
(S_{latt}^{-1}(pa))$ with respect to the quark mass $ma$. It is
expected to be $Z_m Z_{\psi}$ at large $(pa)^2$ from
Eq.~(\ref{tr_S_lat}). These results are very similar to those of
the domain wall fermion~\cite{Tblum1}.

\begin{figure}[tp]
\centering{\epsfig{angle=90,figure=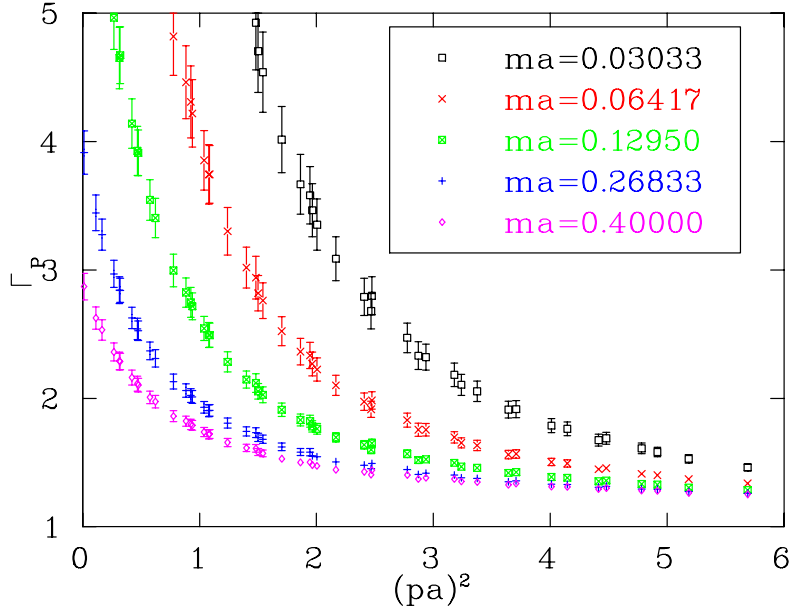,height=8cm} }
\parbox{130mm}{
\caption{$\Gamma_P$ versus $(pa)^2$ with different masses. Here, one can clearly see
the strong divergent behavior for small quark masses. }
\label{figgammap}}
\end{figure}

   From Eqs.~(\ref{Gamma_P}) and (\ref{tr_S_lat}), one finds that
\begin{equation}   \label{gamma_p}
 \Gamma_P (pa,pa) = \frac{Z_{\psi}}{Z_P} - C_1 Z_{\psi} \frac{a^3 \langle \bar{q}q\rangle}
 {ma(pa)^2} + O(1/p^4).
\end{equation}
Since the quark condensate $\langle \bar{q}q\rangle$ has a
contribution of $\frac{\langle |Q|\rangle}{mV}$ from the zero
modes due to the topological charge $Q$, it is expected that
$\Gamma_P (pa,pa)$ has $1/m^2$ and $1/m$ singularities as $m
\rightarrow 0$. Thus, it is suggested~\cite{Tblum1} to fit
$\Gamma_P (pa,pa)$ with the functional form involving pole terms.
As illustrated in Fig.~\ref{figgammap}, the singular behavior in
$m$ is quite visible. It is suggested in Ref.~\cite{Tblum1} to fit
the $\Gamma_P$ with a double and single pole form for each $pa$
%[NOTE: TO BE CONSISTENT WITH THE FINITE MA DISCUSSION LATER, THE FITTING SHOULD
%INCLUDE A $MA^2$ TERM WHICH IS LINEAR IN M, I.E. $\Lambda_{QCD}m a^2$.
%I have tried the fit to include the  $\Lambda_{QCD}m a^2$, but the data show
%there are overfit,  that $\xi^2$ is very small, and the error in fitted value is large.]
\begin{equation}  \label{pole_form}
\Gamma_P (pa,pa) =  \frac{c_{1,P}}{(am)^2} + \frac{c_{2,P}}{(am)} + c_{3,P} +
c_{4,P} (am)^2 \, .
\end{equation}
This is appropriate for the case studied in Ref.~\cite{Tblum1}
where the lattice volume is relatively small (the space-time
volume is $\sim 10\, {\rm fm}^4$) so that the zero mode
contribution is substantial and the quark mass is relatively heavy
so that the quenched chiral log is not significant. In our case,
the space-time volume at $184\, {\rm fm}^4$ is much larger. As
such, the zero mode contribution is expected to be smaller. In our
study of the quenched chiral log in the pion mass~\cite{kent3}
with the same lattice, it is found that the pion mass is basically
the same when calculated from either the $\langle PP\rangle,
\langle A_4 P\rangle, \langle A_4 A_4\rangle$, or $\langle PP
-SS\rangle$ correlators, indicating that the zero mode effects are
small and negligible within statistical errors, even for the
smallest pion mass at $\sim 180$ MeV. Our lowest pion mass is
$212$ MeV in the current study. On the other hand, the quenched
chiral log is quite prominent with pion mass less than 400 MeV.
Therefore, we believe that the more appropriate approach is to
relate the quark condensate in Eq.~(\ref{gamma_p}) to the pion
mass through the Gell-Mann-Oakes-Renner relation
\begin{equation}  \label{GOM}
\langle\bar{q}q\rangle = - {m_{\pi}^2 f_{\pi}^2\over 2m},
\end{equation}
and use the power form for the pion mass where the leading log is
re-summed through the cactus diagrams~\cite{sha92}, i.e.
\begin{eqnarray}   \label{chilog}
m_{\pi}^2(ma) = A\,(ma)^{1\over {1+\delta}} + B(ma)^2.
\end{eqnarray}
Here $\delta$ is the quenched chiral log parameter.

Neglecting higher order terms and expanding $f_{\pi} \simeq f_{\pi}(0) + c\,(ma)$, we can approximate Eq.~(\ref{gamma_p})
as
\begin{eqnarray}   \label{G-P}
\Gamma_{P,\,\hbox{latt}}(ap,ma) &\simeq& A_1
{\left\{(ma)^{{1\over {1+\delta}}-2} + a_P (ma)^{{1\over {1+\delta}}-1}\right\}} + A_2 + C_P(ma^2) + D_P(ma)^2,
\end{eqnarray}
where
\begin{eqnarray}  \label{AaB_P}
A_1 &=&{A a^3 f_{\pi}^2(0) \over 2(pa)^2}\,C_1Z_{\psi}\,,\nonumber\\
a_P &=&{c \over f_{\pi}(0)}\,,\nonumber \\
A_2 &=& {B a^3 f_{\pi}^2(0) \over 2(pa)^2}\,C_1Z_{\psi} + Z_mZ_{\psi}\,.
\end{eqnarray}

We plot $\Gamma_P (pa,pa)$ in Fig.~\ref{figgammap} as a function of $(pa)^2$ for several $ma$. It is clear
that at small $(pa)^2$, it has a singular behavior for small $ma$ which is presumably due to
the divergent terms associated with the quenched chiral $\delta$.

\begin{figure}[tp]
\centering{\epsfig{angle=90,figure=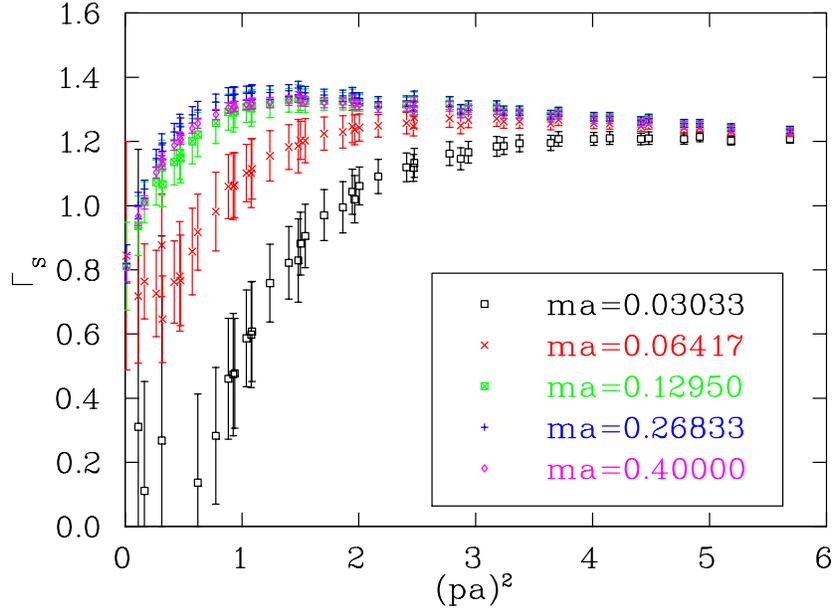,height=8cm} }
\parbox{130mm}{
\caption{$\Gamma_S$ versus $(pa)^2$ with different masses. The chiral log behavior is also
quite visible.}
\label{figgammas}}
\end{figure}

%\begin{figure}[tp]
%\centering{\epsfig{angle=90,figure=./graphs/Vpc124fit_psq.ps,height=8cm} }
%\centering{\epsfig{angle=90,figure=./graphs/Vsc124fit_psq.ps,height=8cm} }
%\parbox{130mm}{
%\caption{ Other coefficients of $\Gamma_P$ and $\Gamma_S$ fitting.
%In both cases, the coefficients $C_{1,S}$ , $C_{1,P}$ and $C_{2,S}$
%are quite small.
%}
%\label{figgammapsc}
%\end{figure}

%In Fig.~\ref{figgammas}, we see non-monotonic behavior of $\Gamma_S$ versus $ma$,
%The data should be in the decreasing order from the top to bottom for the mass if it has
%monotonic behavior,
%actually the data of the heaviest $ma$ lies  below the data of the second heaviest.

From the vector Ward identity, one has the relation
\begin{equation} \label{Gamma_S}
\Gamma_S  = \frac{1}{12} \frac{\partial Tr[S(p)^{-1}]}{\partial m}.
\end{equation}
From Eq.~(\ref{tr_S_lat}), the above $\Gamma_S$  can be
approximated with
\begin{equation}
\Gamma_S (pa,pa)  = \frac{Z_{\psi}}{Z_S} + \frac{C_1 Z_{\psi}}{(pa)^2} \frac{\partial
a^3 \langle \bar{q}q\rangle}{\partial ma} + ...
\end{equation}
for large $(pa)^2$.

Similar to the pseudo-scalar case, substituting the pion mass in Eq.~(\ref{chilog}) and
the Gell-Mann-Oakes-Renner relation in Eq.~(\ref{GOM}) to Eq.~(\ref{Gamma_S}) and neglecting
the higher order terms beyond $(ma)^2$, we obtain
\begin{eqnarray}   \label{G-S}
\Gamma_{S,\,\hbox{latt}}(ap,ma) &=& A_1
{\left\{-\,{\delta\over {1+\delta}}\,
(ma)^{{1\over {1+\delta}}-2} + a_S\,{1\over {1+\delta}}\, (ma)^{{1\over {1+\delta}}-1}\right\}} \nonumber\\
&&\hspace*{1.5in}+ A_2 + C_S(ma^2) + D_S(ma)^2,
\end{eqnarray}

At first glance, it appears that the quantity $Z_mZ_{\psi}$ in $A_2$ is not separable from the other term in
Eq.~(\ref{AaB_P}). However, we should note that we know the values
of $A, B$ and $f_{\pi}(0)$ in Eq.~(\ref{AaB_P}) from an earlier study of the pion mass and decay
constant~\cite{kent3}. After fitting $A_1$ and $A_2$  in Eqs.~(\ref{G-P}) and (\ref{G-S}),
one can compare the first term in $A_2$ with $A_1$ to obtain $Z_mZ_{\psi}$.
As we shall see later, it turns out the first term in $A_2$ is $O(10^{-2}$) times
smaller than $Z_mZ_{\psi}$.

 In order to obtain $Z_mZ_{\psi}$ and assess its finite $m$ behavior, we
first subtract the divergent terms in Eqs.~(\ref{G-P}) and (\ref{G-S}) and then fit the subtracted
vertex functions linear and quadratic in $m$, i.e. with $ma^2$ and $m^2a^2$ terms. In the following
sub-section, we shall detail our fitting methods and give the results. We should note in passing that it has
been suggested to use non-degenerate quark masses to avoid the divergence in $\Gamma_P$~\cite{gv00,GHR}. However, due to
the complication of the quenched chiral log, it is not applicable here.

\subsubsection{Fitting}

We adopt the fitting procedure used to fit the chiral logs in the
pion mass~\cite{kent3} and the Roper resonance in the nucleon
correlator~\cite{mcd05} with priors. From the chiral log fit of
the pion mass~\cite{kent3}, we obtain $\delta$ to be in the range
of $0.20$ -- $0.15$ when the maximum pion mass for the fitting
range is set to be $\sim 500$ MeV -- $900$ MeV. Since we are
fitting $\Gamma_P$ and $\Gamma_S$ in the similar quark mass range,
we put a weak constraint on the value of $\delta$ with $\delta =
0.18(5)$ which covers the range of $\delta$ in the fit of the
quenched chiral log in the pion mass. Data corresponding to the
few lowest masses are first fitted with $A_{1P}(A_{1S}), a_P(a_S)$
and $A_{2P}(A_{2S})$ and then these parameters are constrained
with those fitted values to fit the whole range of the masses with
the forms in Eqs.~(\ref{G-P}) and (\ref{G-S}).
%Fitting is done both by including and excluding correlation between
%different quark masses.
It is observed that these vertex functions are highly correlated between different masses and
that the correlation increases with higher momentum. This, we believe, is due to the fact that the quark
masses that we are interested in are all much smaller than the external momentum of $pa = 4.145$ which
corresponds to $\mu$ = 2 GeV that we will use to eventually match to the $\overline{\rm MS}$ scheme at this
scale. In this sense, the high correlation is a generic feature that this non-perturbative
renormalization procedure faces for light quarks.

\begin{figure}[ht]
\includegraphics[height=7.0cm]{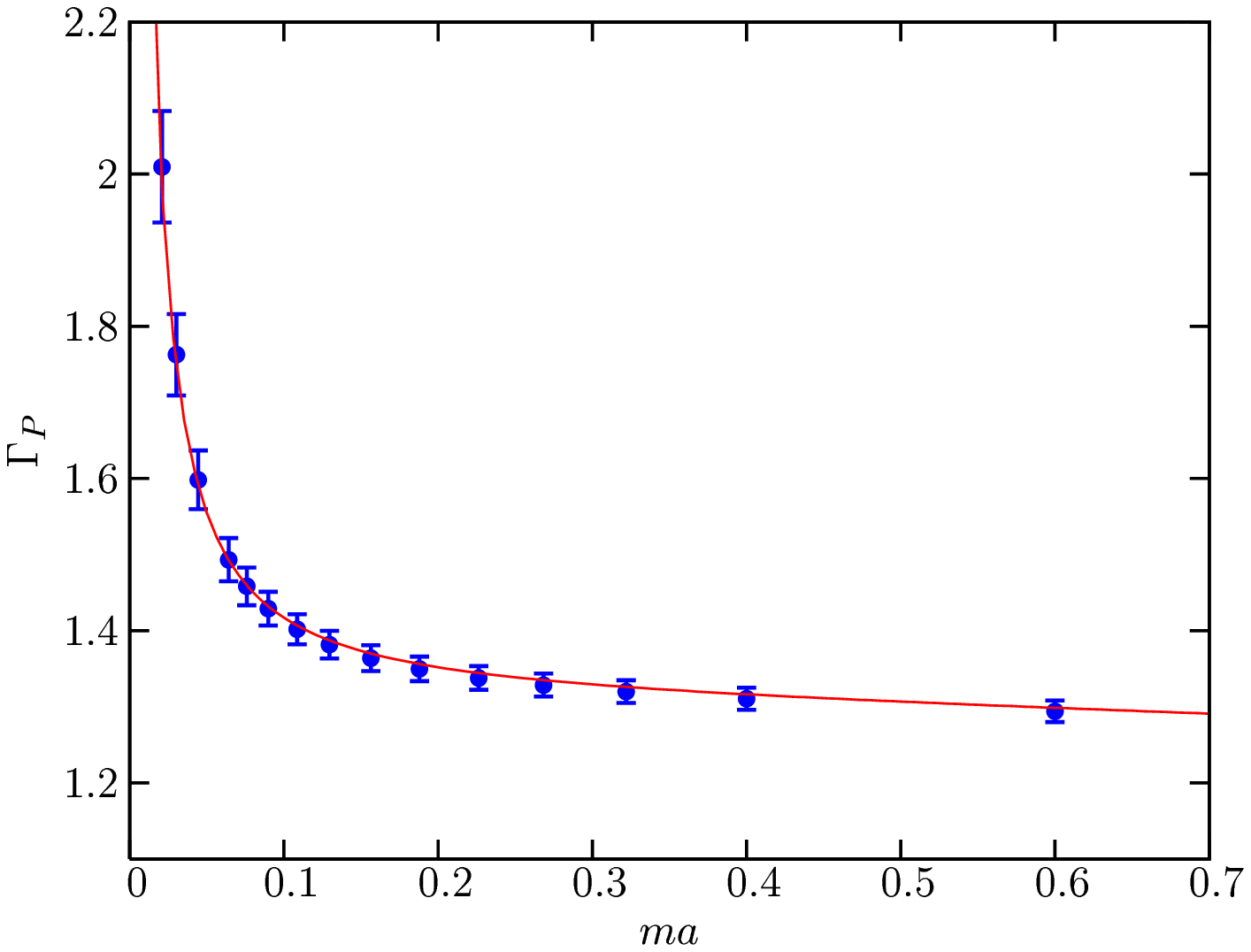}
\includegraphics[height=7.0cm]{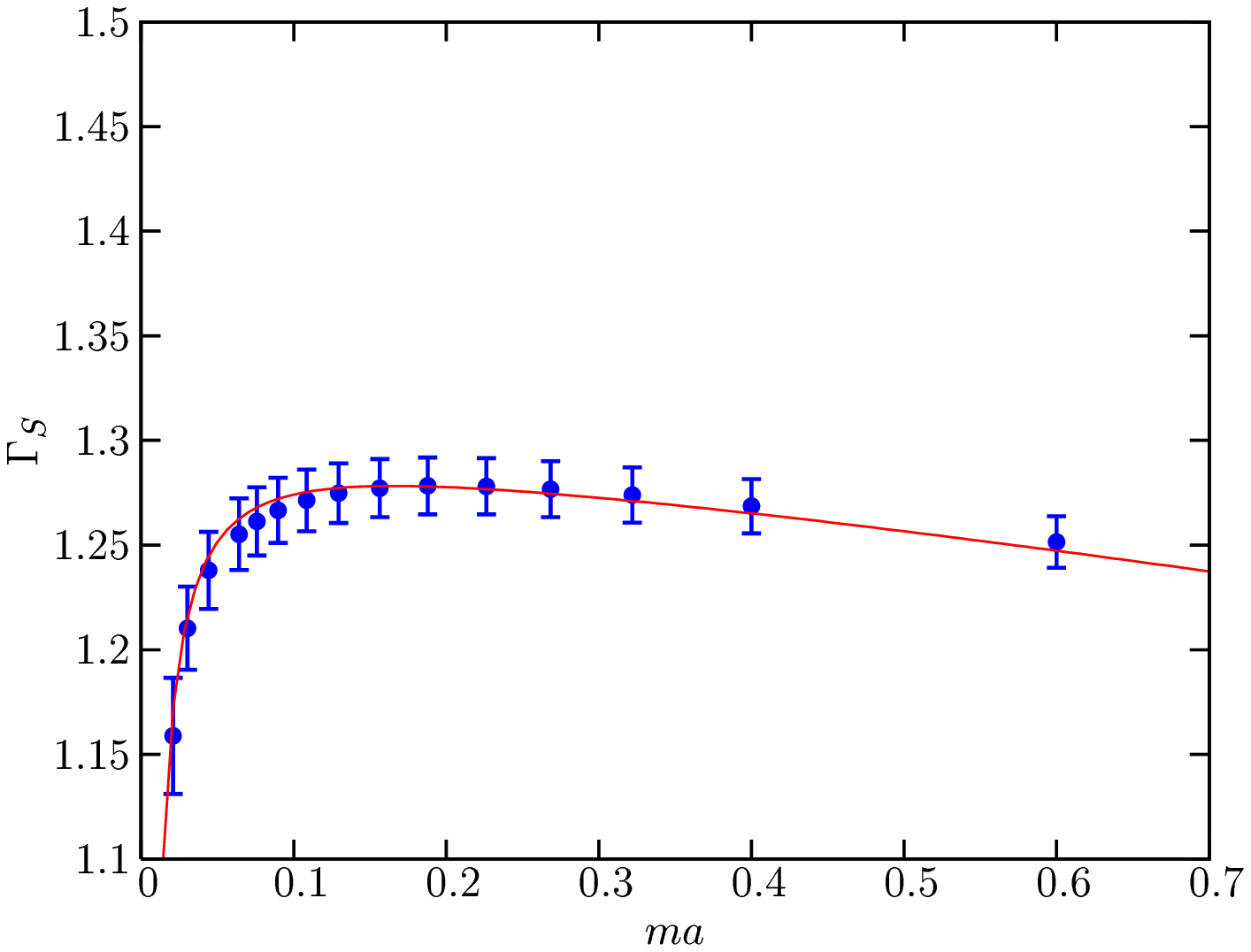}
\caption{Vertex functions for scalar (top) and pseudo-scalar (bottom)
channels at the momentum corresponding to $\mu$ = 2 GeV.}
\label{GPGS1}
\end{figure}

\begin{center}
\begin{table}
\caption{Fitted parameters corresponding to Eqs.~(\ref{G-P}) and (\ref{G-S})}
\vspace*{0.1in}
\begin{tabular}{cccccccc}
\hline
\hline
Channel & $A_1$ & $a_{P}(a_S)$ & $\delta$ & $A_2$ & $C$ & $D$ & $\chi^2$/dof\\
\hline
P & 0.008(1)&0.08(9)&0.163(18)&1.305(15)&$-0.024(6)$&$-0.022(5)$&1.29\\
S & 0.011(5)&0.14(42)&0.171(24)&1.302(17)&$-0.079(8)$&$-0.020(5)$&1.47\\
\hline
\hline
\end{tabular}
\label{table:GPGS}
\end{table}
\end{center}

We show in Fig.~\ref{GPGS1} the vertex functions $\Gamma_P(pa,ma)$
and $\Gamma_S(pa,ma)$ at $pa = 4.145$ as a function of $ma$,
together with the fitted curves based on Eqs.~(\ref{G-P}) and
(\ref{G-S}). The fitted parameters are given in
Table~\ref{table:GPGS}. We see from the fitted parameters $A_1,
a_P (a_S), \delta$ and $A_2$ from $\Gamma_P$ agree with those
fitted from $\Gamma_S$ respectively, as we expected. This supports
our supposition that the singular behaviors in both the
$\Gamma_P(pa,ma)$ and $\Gamma_S(pa,ma)$ are due to the quenched
chiral log in $\langle \bar{q}q\rangle$. We also tried to fit the
pseudoscalar vertex function in the form in Eq.~(\ref{pole_form})
and found that it does not fit well --- the $\chi^2$ is too large.

We note that the fitted central values of $\delta$ tend to be on the low side
compared to our previous fit of the pion mass which gives $\delta = 0.20(3)$~\cite{kent3} and is in
agreement with $\delta \sim 0.23$ as deduced from the topological susceptibility calculation with the
overlap operator~\cite{dgp05}. We think the reason is that the smallest quark mass in the present fit which corresponds
to $m_{\pi} \sim 250$ MeV is larger than that in the previous fitting of the pion mass
which corresponds to  $m_{\pi} \sim 180$ MeV. According to the detailed study~\cite{kent3} of the quenched chiral
log as a function of the fitting range of quark masses, this behavior of a smaller $\delta$ for a  higher quark mass
range is to be expected.

 When we take the value of $A = 1.3$ and $B = 1.1$ from our pion mass chiral log fit
(Fig.~12 in Ref.~\cite{kent3}) in the relevant mass range, it follows from Eq.~(\ref{AaB_P}) that
\begin{eqnarray}
R= {Ba^3 f_{\pi}^2(0)\over 2\,(pa)^2}\,C_1Z_{\psi} = \frac{B A_1}{A} = 0.0068,
\end{eqnarray}
which is two orders of magnitude smaller than $A_2$. We shall subtract this contribution from $A_2$ in
Eq. (\ref{AaB_P}) to obtain $Z_mZ_{\psi}$ which changes its value by about half a $\sigma$ which is not significant.

To eventually obtain $Z_S$ and $Z_P$ and their respective finite $m$ dependence, we
define the subtracted $\Gamma_P$ and $\Gamma_S$ by taking out the divergent terms in Eqs.~(\ref{G-P})
and (\ref{G-S}) and the first term in $A_2$ in Eq.~(\ref{AaB_P}) on each Jackknife sample ($J$) as
\begin{eqnarray}  \label{PS_sub}
\Gamma_{P}^{sub,J}(ma) &=& \Gamma_{P}^{J}(ma) -  A_{1J}^{P}
{\left\{(ma)^{{1\over {1+\delta_J}}-2} + a_{J}^{P}\,(ma)^{{1\over {1+\delta_J}}-1}\right\}} -\frac{B\,A_{1J}^P}{A}\,,\\
\Gamma_{S}^{sub,J}(ma) &=& \Gamma_{S}^{J}(ma) -  A_{1J}^{S}
{\left\{-\,{\delta_J\over {1+\delta_J}}\,
%(ma)^{{1\over {1+\delta_J}}-2} + a_{J}^{S}\,{1\over {1+\delta_J}}\,
(ma)^{{1\over {1+\delta_J}}-2} + {a_{J}^{S}\over {1+\delta_J}}\,
(ma)^{{1\over {1+\delta_J}}-1}\right\}} - \frac{B\,A_{1J}^S}{A}\,.
\end{eqnarray}

%\begin{figure}[tp]
%\centering{\epsfig{angle=90,figure=./graphs/Vpfit_psq.ps,height=8cm} }
%\centering{\epsfig{angle=90,figure=./graphs/Vsfit_psq.ps,height=8cm} }
%\parbox{130mm}{
%\caption{Subtracted $\Gamma_P^{sub}$ and $\Gamma_^{sub}$ at $ma=0$ as a function of $(pa)^2$.}
%\label{figgammaps}
%\end{figure}

\begin{figure}[!ht]
\centering{\epsfig{angle=90,figure=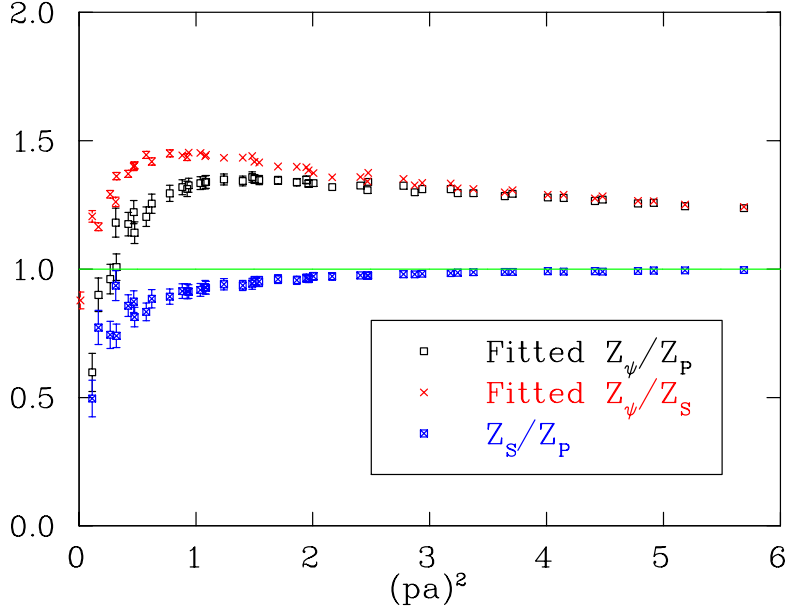,height=8cm} }
\caption{ $Z_{\psi}/Z_P$ and  $Z_{\psi}/Z_S$ and the ratio $Z_P/Z_S$ as a function of $(pa)^2$. }
\label{checksp}
\end{figure}

%Question for Jianbo: what are plotted in Fig,9? They seem different from those
%in Fig. 10. Shouldn't they be the same?
% Yes, I have fixed it.

From Eqs.~(\ref{G-P}), (\ref{AaB_P}), (\ref{G-S}), and
(\ref{PS_sub}), we see that the ratios of $Z_{\psi}/Z_P$ and
$Z_{\psi}/Z_S$ are the subtracted vertices $\Gamma_P$ and
$\Gamma_S$ at the massless limit for each $pa$ in the RI scheme,
i.e.
\begin{equation}
\frac{Z_{\psi}(pa)}{Z_{P,S}(pa)} = \Gamma_{P,S}^{sub}(pa, ma =0).
\end{equation}

We plot in Fig.~\ref{checksp}, $Z_{\psi}/Z_P$, $Z_{\psi}/Z_S$, and the ratio $Z_S/Z_P$ as a
function of $(pa)^2$. We see that for $(pa)^2 > 3$, the ratio goes to unity which is a confirmation
that our fitting procedure does not spoil the expected chiral property $Z_P = Z_S$ for the overlap fermion.

\subsection{The tensor current}

\begin{figure}[!hb]
\centering{\epsfig{angle=90,figure=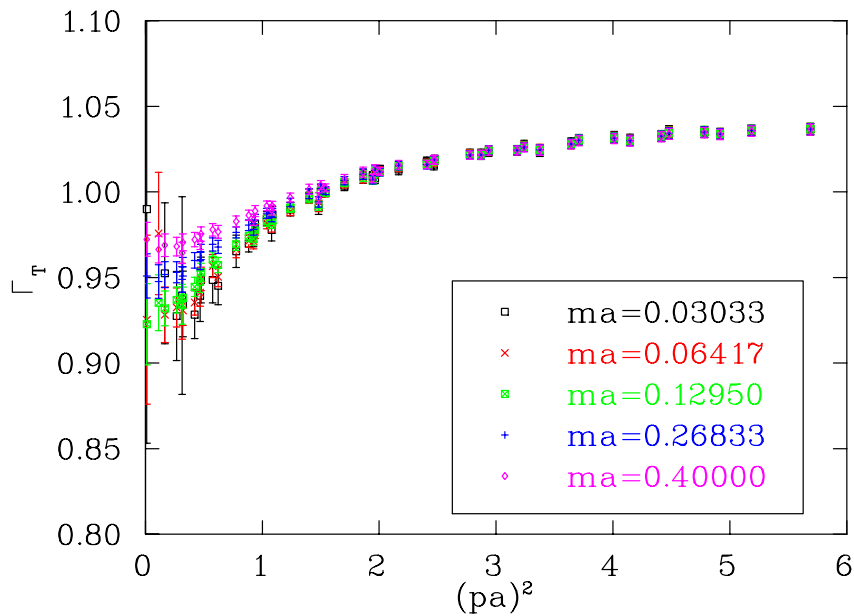,height=8cm} }
%\parbox{130mm}{
\caption{$\Gamma_T$ versus $(pa)^2$ with different masses.
It has no significant mass dependence }
\label{figgammt}
\end{figure}

In Fig.~\ref{figgammt}, we show $\Gamma_T$
versus $(pa)^2$ with different masses.  We can see that at moderate to large
$(pa)^2$, $\Gamma_T$ is not sensitive to the quark masses. The chiral limit value
is obtained by a linear plus quadratic fitting as for the case of vector and axial vector currents
and the results will be presented in the next section.

\section{Running of the renormalization constants} \label{sec:running}

  In general, one can choose to define the renormalization conditions for different
$p$ and $p^\prime$ in Eq.~(\ref{lambda_pq}). But the virtuality of the quark states must be much larger than
$\Lambda_{QCD}$. This is so because in order to obtain physical results at certain
scale (e.g. momentum or mass), one needs to combine the matrix element of the
renormalized operator $O(\mu)$ with the Wilson
coefficient function. The latter is usually computed in continuum perturbation theory
by expanding in $\alpha_s^{\overline{\rm MS}}$ at a scale of order of $\mu$. Thus, for
the validity of perturbation calculation, $\mu$ must be large. On the other hand, one would
like to have $\mu \ll 1/a$ in order to have smaller $O(a)$ effects. When spontaneous
symmetry breaking takes place, as is in QCD, a large $\mu$ may not be enough due
to the presence of the Goldstone boson. For example, at low momentum transfer
$q = p' - p$, the Green's function can have a Goldstone boson pole like  $1/q^2$.
However for fermions (but not for a scalar particle), this contribution will be $1/p^2 = 1/\mu^2$
smaller than the perturbative contribution even when $q^2 = 0$ ~\cite{NPM}.
Thus, it is desirable to have a window $\Lambda_{QCD} \ll \mu \ll 1/a$
so that both the non-perturbative effects and the lattice artifacts are small.
In practice, one finds that the renormalization procedure prescribed here works well
for $(\mu a)^2$ as large as 6.  In the current case, this corresponds to $\mu \sim 2.5$ GeV.

The renormalized operators are  defined as
\begin{equation}
Z_{O} O_{\rm bare} = O_{\rm ren} \, .
\label{Zvalue}
\end{equation}
The fact that the bare operator is independent of the renormalization
scale $\mu^2$ gives the renormalization group (RG) equation,
\begin{eqnarray}
\mu^2 \frac{d}{d \mu^2} O_{\rm ren}
&=&  \frac{1}{Z_O} \mu^2 \frac{d Z_{O}}{d \mu^2}  O_{\rm ren} \,
\\
&=&
-\frac{\gamma_O}{2} O_{\rm ren} \, .
\end{eqnarray}
where
\begin{equation}
\gamma_O = -\frac{2\mu^2}{Z_O}\frac{d Z_{O}}{d \mu^2},
\end{equation}
is the anomalous dimension.

The solution of $Z_O(\mu^2)$ can be written in the following
form
\begin{equation}   \label{eq:evo}
Z_O\left(\mu^2\right) =
\frac{
C_O\left({\mu}^2\right)
}{
C_O\left({\mu'}^2
\right)
} Z_O\left({\mu'}^2\right) \, .
\end{equation}
Expanding the anomalous dimension in the coupling constant $\alpha_s$
\begin{eqnarray}
\gamma_O &=& \sum_i \gamma^{(i)}_O \left( \frac{\alpha_s}{4 \pi}\right)^{i+1} \, ,
\end{eqnarray}
and considering the
running of $\alpha_s$ in QCD $\beta$ function $\beta(\alpha_s)$ perturbatively
\begin{eqnarray}
\frac{\beta (\alpha_s)}{4\pi} & =& \mu^2\frac{d}{d\mu ^2} \left( \frac{\alpha_s}{4\pi} \right) \,
= \, - \sum_{i=0}^\infty \beta _i\left( \frac{\alpha_s}{4\pi } \right)^{i+2} \, ,
\end{eqnarray}
where $\beta_i$ are the coefficients of the QCD $\beta$ function $\beta(\alpha_s)$, one can solve
for the coefficient functions $C_O$ perturbatively.

The four loop solution~\cite{Chetyrkin:20004l} of the coefficient function in Eq.~(\ref{eq:evo})
is  (we have suppressed the subscripts for the specific operator $O$)
\begin{eqnarray}
{C(\mu^2)} &=&
\left(\frac{\alpha_s(\mu)}{\pi}\right)^{\bar{\gamma_0}}  \left\{ 1 +
\left(\frac{\alpha_s(\mu)}{4 \pi}\right) (\bar{\gamma_1} - \bar{\beta_1}\bar{\gamma_0})
\right.
\nonumber \\
&+&
\frac{1}{2} \left(\frac{\alpha_s(\mu)}{4 \pi}\right)^2
\left[
(\bar{\gamma_1} - \bar{\beta_1}\bar{\gamma_0})^2
+
\bar{\gamma_2} + \bar{\beta_1}^2\bar{\gamma_0}
- \bar{\beta_1}\bar{\gamma_1} -\bar{\beta_2}\bar{\gamma_0}
\right]
\nonumber \\
&+&   \left(\frac{\alpha_s(\mu)}{4 \pi}\right)^3
\left[
\frac{1}{6}(\bar{\gamma_1} - \bar{\beta_1}\bar{\gamma_0})^3
+
\frac{1}{2}(\bar{\gamma_1} - \bar{\beta_1}\bar{\gamma_0})
(
\bar{\gamma_2} + \bar{\beta_1}^2\bar{\gamma_0}
- \bar{\beta_1}\bar{\gamma_1} -\bar{\beta_2}\bar{\gamma_0}
)
\right.
\nonumber \\
&&
+\frac{1}{3}\left(
\left.\left.
\bar{\gamma_3}
-\bar{\beta_1}^3\bar{\gamma_0} + 2\bar{\beta_1} \bar{\beta_2}\bar{\gamma_0}
-\bar{\beta_3}\bar{\gamma_0} + \bar{\beta_1}^2\bar{\gamma_1}
- \bar{\beta_2}\bar{\gamma_1} - \bar{\beta_1}\bar{\gamma_2}
\right)
\right]
\right\}
{},
 \label{c(x)}
\end{eqnarray}
where
\begin{eqnarray}
\overline{\gamma}_{i} = \frac{\gamma^{(i)}}{2 \beta_0} \, , \,\,
\,\, \overline{\beta}_i = \frac{\beta_i}{\beta_0} \, .
\end{eqnarray}

The following tables give the anomalous dimensions $\gamma^{(i)}$
for $Z_{\psi}$, $Z_m$ and $Z_T$ in the RI/MOM scheme for the
quenched approximation~\cite{Franco:1998bm,
Chetyrkin:20004l,Gracey}. In the case of chiral fermions, $Z_S$ =
$Z_P$ = 1/$Z_m$, so that $C_S (\mu^2)= C_P (\mu^2) =
1/C_m(\mu^2)$. Note that, in Refs.~\cite{Franco:1998bm,
Chetyrkin:20004l,Gracey}, the definition of $Z$'s is the inverse
of our definition in Eq.~(\ref{Zvalue}).

\begin{table}[!ht]
\begin{center}
\caption{Quenched $Z_\psi$ Anomalous Dimensions $\gamma_\psi^{(i)}$}
\label{tab:zqr}
\bigskip
\begin{ruledtabular}
\begin{tabular}{cccc}
 $\gamma^{(0)}$ & $\gamma^{(1)}$ & $\gamma^{(2)}$  &  $\gamma^{(3)}$\\
\hline
   0 &  44.6667  & 2177.0737 &  130760.2969 \\
\end{tabular}
\end{ruledtabular}
\end{center}
\end{table}

\begin{table}[!ht]
\begin{center}
\caption{Quenched $Z_m$ Anomalous Dimensions $\gamma_m^{(i)}$}
\label{tab:zqs}
\begin{ruledtabular}
\begin{tabular}{cccc}
 $\gamma^{(0)}$ & $\gamma^{(1)}$ &  $\gamma^{(2)}$ &  $\gamma^{(3)}$    \\
\hline
 8  & 252  & 11769.5469 &  557837.9375   \\
\end{tabular}
\end{ruledtabular}
\end{center}
\end{table}

\begin{table}[!ht]
\begin{center}
\caption{Quenched $Z_T$ Anomalous Dimensions $\gamma_T^{(i)}$}
\label{tab:zqt}
\begin{ruledtabular}
\begin{tabular}{ccc}
$\gamma^{(0)}$ & $\gamma^{(1)}$ &  $\gamma^{(2)}$ \\
\hline
2.66667 &  80.4444 & 3268.2996  \\
\end{tabular}
\end{ruledtabular}
\end{center}
\end{table}

The coupling constant itself is running with respect to
$\mu$. The four loop formula is given by~\cite{beta4loop}
\begin{eqnarray}
\frac{\alpha_s}{4 \pi}
&=&
\frac{1}{\beta_0 \ln \left( \mu^2 / \Lambda^2_{QCD} \right) }
-
\frac{\beta_1 \ln \ln \left( \mu^2 / \Lambda^2_{QCD} \right)
}{
\beta_0^3 \ln^2 \left( \mu^2 / \Lambda^2_{QCD} \right)}
\\ \nonumber
&&
+
\frac{1}{\beta_0^5 \ln ^3\left( \mu^2 / \Lambda^2_{QCD} \right)
}\left\{
\beta_1^2 \ln^2 \ln \left( \mu^2 / \Lambda^2_{QCD} \right)
- \beta_1^2 \ln \ln \left( \mu^2 / \Lambda^2_{QCD} \right)
\right.
\\ \nonumber
&&
\left.
+ \beta_2 \beta_0 - \beta_1^2
\right\}
+ \frac{1}{\beta_0^7 \ln ^4\left( \mu^2 / \Lambda^2_{QCD} \right) }
\left\{ \beta_1^3 \ln^3 \ln \left( \mu^2 / \Lambda^2_{QCD} \right) \right. \\ \nonumber
&&
\left.
-\frac{5}{2} \beta_1^3 \ln^2 \ln \left( \mu^2 / \Lambda^2_{QCD} \right)
- \left( 2 \beta_1^3 - 3 \beta_0 \beta_1 \beta_2 \right) \ln \ln \left( \mu^2 / \Lambda^2_{QCD} \right)
\right. \\ \nonumber
&&
\left.
+ \frac{1}{2} \left( \beta_1^3 - 3 \beta_0^2 \beta_3 \right) \right\}
\end{eqnarray}

The QCD $\beta$-function is scheme independent only up to two
loops. The additional terms of the expansion have been computed in
the $\overline{\rm MS}$ scheme in Ref.~\cite{b4}.

In this work, the value of $\alpha_s$ was calculated at four loops using
a lattice value of $\Lambda_{\rm QCD}$
taken from Ref.~\cite{Capitani:1998mq}
as
\be \Lambda_{\rm QCD} = 238 \pm 19 ~\mathrm{MeV} \, .  \ee

\begin{figure}[hb]
\centering{\epsfig{angle=90,figure=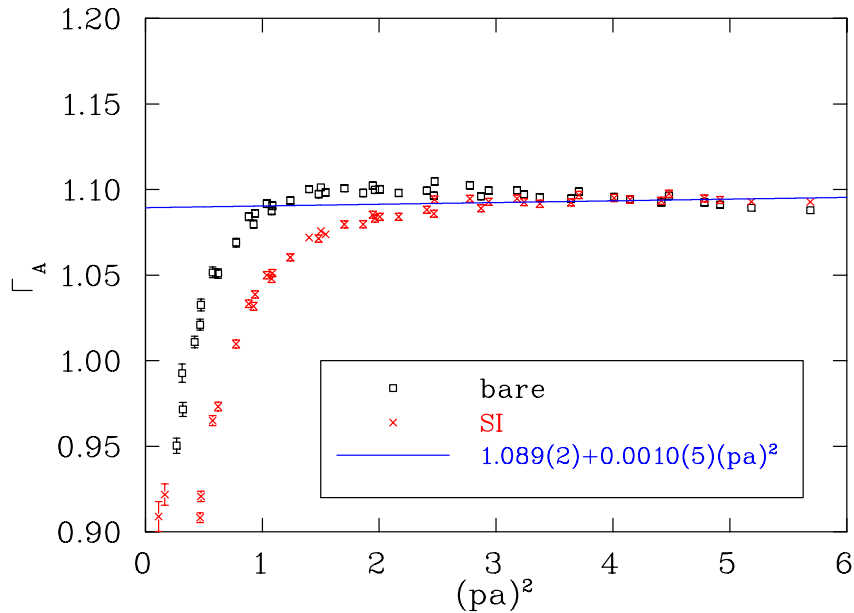,height=8cm} }
%\parbox{130mm}{
\caption{$\Gamma_A$ (labelled as ``bare'') and the scale invariant
$\Gamma_A^{\rm SI}$ versus $(pa)^2$ in the chiral limit. They
coincide near $(pa)^2=4.1$. The later is almost $(pa)^2$
independent after $(pa)^2 > 2.4$, the slope versus $(pa)^2$ is
about 0.001.} \label{figvasi}
\end{figure}

\begin{figure}[h]
\centering{\epsfig{angle=90,figure=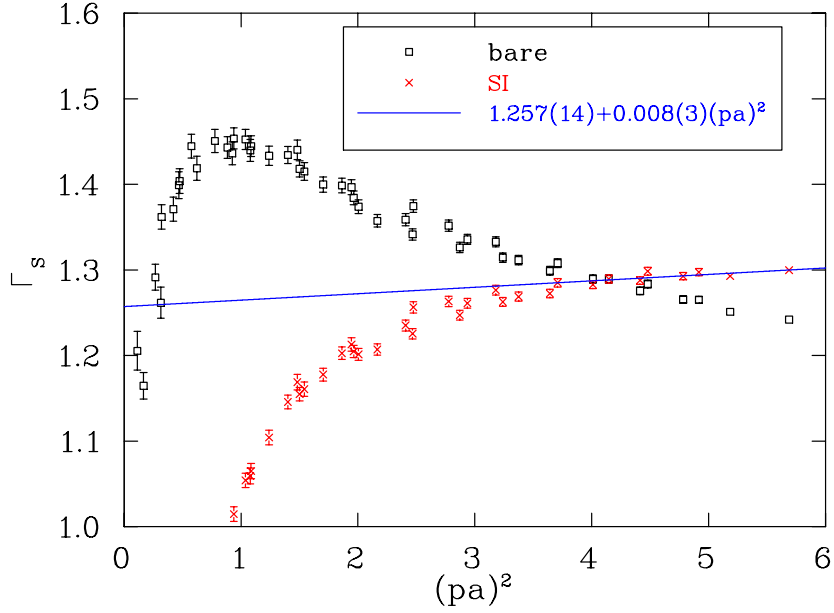,height=8cm} }
%\parbox{130mm}{
\caption{ The same as Fig.~(\ref{figvasi}) for $\Gamma^{\rm
sub}_S$ and $\Gamma_S^{\rm sub,SI}$. The slope of SI versus
$(pa)^2$ is about 0.008 beyond $(pa)^2$  $\ge$ 4.0. }
\label{figvssi}
\end{figure}

\begin{figure}[h]
\centering{\epsfig{angle=90,figure=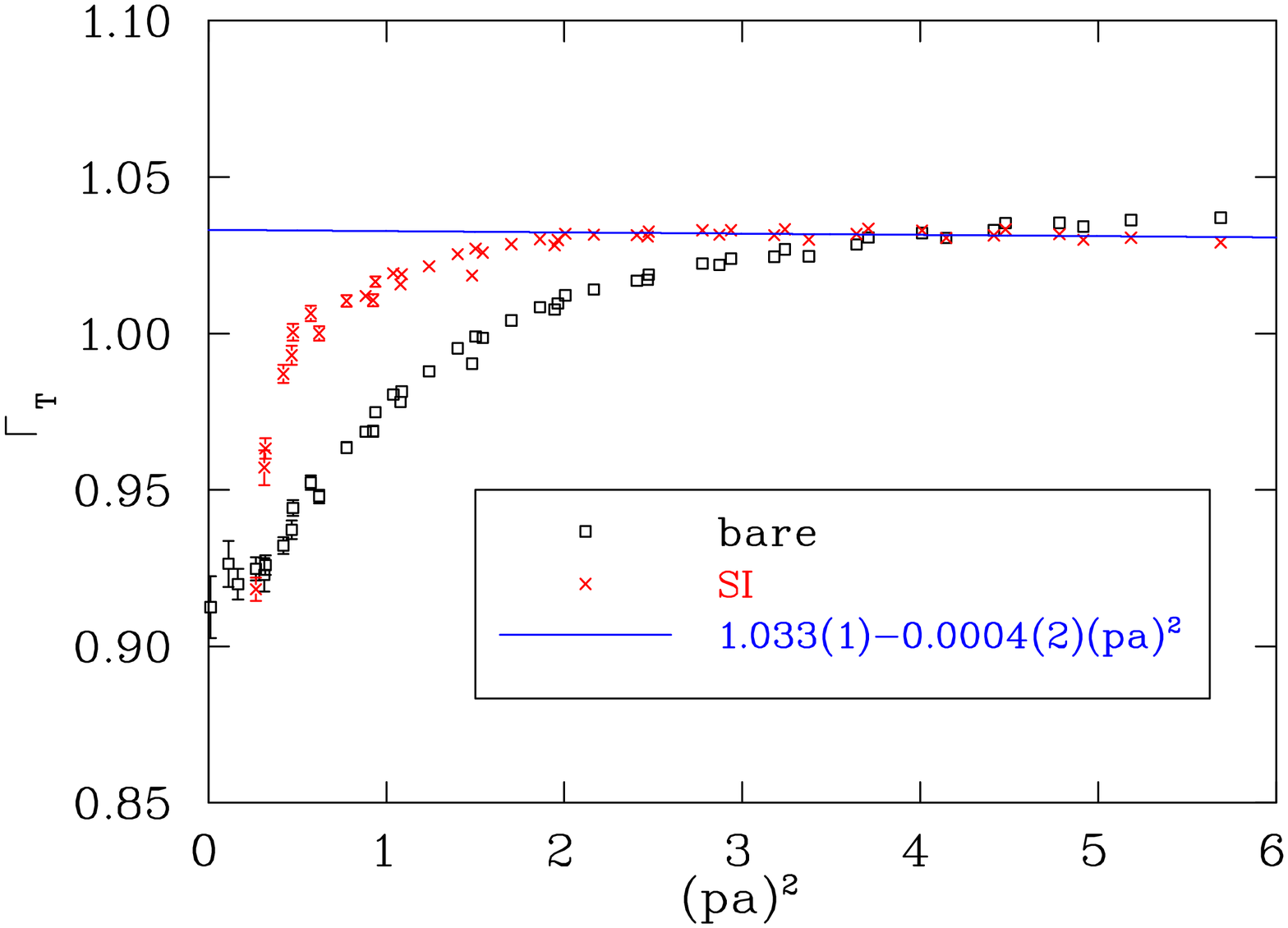,height=8cm} }
%\parbox{130mm}{
\caption{The same as Fig.~(\ref{figvasi}) for $\Gamma_T$ and $\Gamma_T^{\rm SI}$. The later  is almost
$(pa)^2$ independent after $(pa)^2 > 2.0$, the slope versus $(pa)^2$ is about -0.0004. }
\label{figvtsi1}
\end{figure}

Both $Z_A$ and $Z_V$ are scale independent, but this is not the
case for $Z_\psi$. The scale invariant (SI) vertex for the axial
and vector current is defined by removing the renormalization
group running of $Z_{\psi}$ as \be \Gamma_{A,V}^{\rm
SI}\left((ap)^2\right) =
\Gamma_{A,V}\left((ap)^2\right)/C_{\psi}\left((ap)^2\right),
\label{eq:running} \ee where $C_{\psi}$ is defined in
Eq.~(\ref{c(x)}) with the anomalous dimension coefficients from
Table~\ref{tab:zqr}. We normalize $C_{\psi}((\mu a)^2 = 4.15)=1$,
which corresponds to $\mu$ = 2.0 GeV, in order to compare with the
$(pa)^2$ dependence of $\Gamma_{A,V}\left((ap)^2\right)$.

Fig.~\ref{figvasi} shows both $\Gamma_A\left((ap)^2, ma = 0\right) $ and
$\Gamma_A^{\rm SI}\left((ap)^2, ma=0\right)$ as a function of $(pa)^2$. By comparing them, we see
that the renormalization group running due to $Z_\psi$ is not
appreciable for $(pa)^2 > 3$, but it does tend to make the SI data flatter as a function of $(pa)^2$.
The remaining scale dependence of $\Gamma_A^{\rm SI}\left((ap)^2, ma=0\right)$
is very small. A plausible explanation for it is an
$(ap)^2$ error~\cite{Tblum1}. Fitting the remaining scale dependence to
the form~\cite{Tblum1}
\be
\Gamma_A^{\rm SI}\left((ap)^2\right) = \Gamma_A^{\rm SI} + c  (ap)^2 \, ,
\label{sifit}
\ee
for a range of momenta that is chosen to be ``above'' the region where
the condensate effects are important, one can obtain the
scale invariant $\Gamma_A$ which is denoted as $\Gamma_A^{\rm SI}$.

When a linear fit of the SI data versus $(ap)^2$ is performed for
$2.4 < (ap)^2 < 5.7$, the gradient is $\approx 0.001$. In the
ideal case, the gradient should be zero. This small value is thus
interpreted as an $ O(a^2) $ error. This shows that the $(pa)^2$
error of the ratio of $Z_{\psi}$ and $Z_A$ is small, but we don't
know their individual $(pa)^2$ error separately. It appears that
the $(pa)^2$ error in $Z_{\psi}'$ as defined from the quark
propagator in Eq.~(\ref{wf_ren}) is as large as $\sim 10\%$ at $p
= 2$ GeV~\cite{Tblum1,zll04}. However, this relatively large
$(pa)^2$ error in $Z_{\psi}^{\rm SI}$ must be cancelled to a large
extent by that of $Z_A$, resulting in a small $(pa)^2$ error in
$\Gamma_A^{\rm SI}$ (only 0.4\% at $(pa)^2 = 4.1$). As will be
discussed in the next sub-section, we will use $Z_A$ determined
from the chiral Ward identity to obtain $Z_{\psi}$. Since $Z_A$,
in this case, is determined from the pion state at rest, it is at
a momentum scale of $\Lambda_{QCD}$. Thus, it should have small
$(pa)^2$ error. Using this value of $Z_A$ and $\Gamma_A^{\rm
SI}\left((ap)^2, ma=0\right)$ to obtain $Z_{\psi}^{\rm SI}$, and
thus $Z_{\psi}^{RI}(2~{\rm GeV})$ should give a $(pa)^2$ error of
$\sim 0.4\%$ which we shall consider as the systematic error in
$O(a^2)$.

In the case of $\Gamma_S^{\rm sub} = Z_{\psi}/Z_S$, both
$Z_{\psi}$ and $Z_S$ run with $\mu^2$. Fig.~\ref{figvssi} shows
$\Gamma_S^{\rm sub}$ and the corresponding scale invariant (SI)
vertex, $\Gamma_S^{\rm sub,SI}$, after three loop running. We see
that $\Gamma_S^{\rm sub, SI}$ is much flatter than $\Gamma_S^{\rm
sub}$ for $(pa)^2 > 3.0$. The linear fit of the SI data versus
$(pa)^2$ in the range of $4.0 < (ap)^2 < 5.7$, gives a gradient of
0.008(3). This is an order of magnitude larger than that of the
axial vector (and that of the tensor current below). This
relatively larger gradient could be due to the systematic
uncertainty in subtracting the chiral divergence in $\Gamma_S$ or
the mismatch of the $(pa)^2$ errors between $Z_{\psi}^{\rm SI}$
and $Z_S^{\rm SI}$ or both.

The SI result of $\Gamma_T$ which comes from the three loop running of $Z_T$~\cite{Gracey}
and four loop running of $Z_\psi$ is plotted in Fig.~\ref{figvtsi1} along with $\Gamma_T$.
%The "bare" data is taken by a simple linear extrapolation to the chiral limit.
The linear fit to the SI data in the range of  $2.0 < (ap)^2 < 5.7$ gives a gradient of -0.0004(2).

It is worthwhile pointing out that comparing to the Domain Wall fermion case on a
$16^3 \times 32 \times 16$ lattice with Wilson gauge action at $\beta = 6.0$~\cite{Tblum1}, we find
that the remaining scale dependence as measured by the gradient in $(pa)^2$ is comparable
for the $\Gamma_S^{\rm SI}$ case. But the gradient for $\Gamma_A^{\rm SI}/\Gamma_T^{\rm SI}$ is one/two orders of
magnitude smaller than that in the Domain Wall fermion case. Since the Domain Wall fermion with
finite 5th dimension and the rational polynomial approximation of the sign function that we adopt in
the present work are two different approximations of the same overlap fermion in 4 dimensions~\cite{bno04},
the $(pa)^2$ errors in the scale invariant vertices serve as a measure of the
$O(a^2)$ errors of the approximation.

\subsection{Determining $Z_A$ from Ward identity} \label{sec:ZA}

   Before applying the renormalization group running (an inverse operation of
Eq.~(\ref{eq:running})) to match to the $\overline{\rm MS}$ scheme
at certain scale, we need to input  $Z_A$ to Eq. (\ref{3rd}) in
order to determine other renormalization constants from the
respective vertex functions. As explained in Sec.~\ref{RI/MOM}, we
prefer using $Z_A$ from the chiral Ward identity to determine
$Z_{\psi}$ than directly obtaining it from the quark propagator.
This is partly due to the fact that there is ambiguity in the
lattice definition of momentum~\cite{Tblum1} in
Eq.~(\ref{wf_ren}). Furthermore, it is shown in the study with
Domain Wall fermion~\cite{Tblum1} and an earlier study of the
overlap fermion~\cite{zll04} that the $(pa)^2$ errors in the scale
invariant $Z_{\psi}^{\rm SI}$ are quite large.

The renormalization constant $Z_A$ for the axial current
$A_{\mu}=\bar{\psi}(i\gamma_{\mu}\gamma_5(1 - D/2\rho)\frac{\tau^a}{2})\psi$
can be obtained directly through the axial Ward identity
\begin{equation}  \label{awi}
Z_A\partial_{\mu} A_{\mu} = 2 Z_m m Z_P P,
\end{equation}
where $P =\bar{\psi}(i\gamma_5(1 - D/2\rho)\frac{\tau^a}{2})\psi$
is the pseudo-scalar density. For the case of the overlap
fermion~\cite{kent1,kent2,GHR}, $Z_m = Z_S^{-1}$ and $Z_S = Z_P$.
Thus, $Z_m$ and $Z_P$ cancel in Eq.~(\ref{awi}) and one can
determine $Z_A$ to $O(a^2)$ non-perturbatively from the axial Ward
identity using the bare mass $m$ and bare operator $P$.  To obtain
$Z_A$, we shall consider the on-shell matrix elements between the
vacuum and the zero-momentum pion state for the axial Ward
identity
\begin{equation}
Z_A \langle 0|\partial_{\mu} A_{\mu}|\pi(\vec{p} = 0)\rangle
= 2 m \langle 0|P|\pi(\vec{p} = 0)\rangle,
\end{equation}
where the matrix elements can be obtained from the zero-momentum correlators
\begin{eqnarray}
G_{\partial_4A_4P}(\vec{p}= 0, t)& =& \langle\sum_{\vec{x}}\partial_4 A_4(x)
P(0)\rangle \nonumber \\
G_{PP}(\vec{p}= 0, t) & =& \langle\sum_{\vec{x}} P(x) P(0)\rangle.
\end{eqnarray}
The non-perturbative $Z_A$ is then
\begin{equation}   \label{ZA}
Z_A = \lim_{m \rightarrow 0, t \rightarrow \infty} \frac{2 m G_{PP}(\vec{p}= 0, t)}
{G_{\partial_4A_4P}(\vec{p}= 0, t)}.
\end{equation}

\begin{figure}[t]
\vspace*{-6cm}
\centering{\epsfig{figure=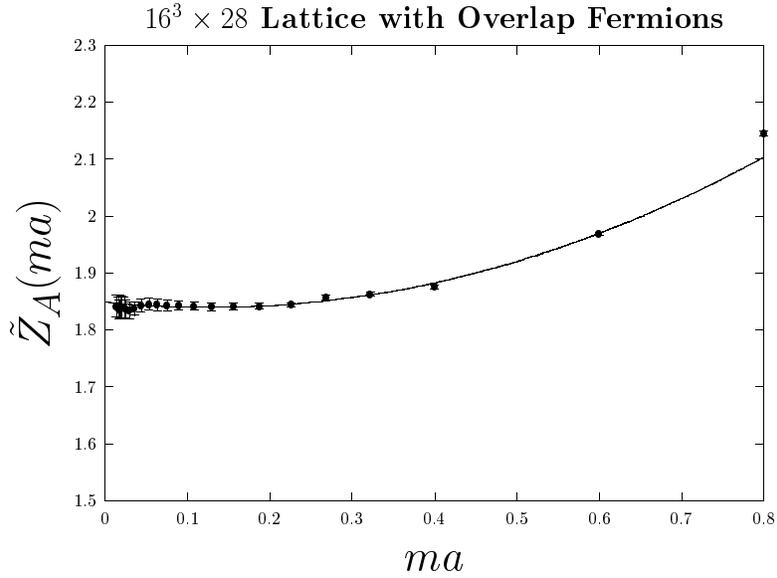,
height=20cm}} \vspace*{-7cm} \caption{$\tilde{Z}_A(ma) $ vs quark
mass $ma$.} \label{Zaawi}
\end{figure}

 Given that the time derivative itself in
$G_{\partial_4A_4P}(\vec{p}= 0, t)$ invokes an $O(a^2)$ error, it would be
better to adopt a definition for $Z_A$ which is devoid of this superfluous
$O(a^2)$ error. This can be achieved by noticing that, at large $t$ where the
pion state dominates the propagator $G_{\partial_4A_4P}(\vec{p}= 0, t)$, one
can effectively make the substitution
\begin{equation}   \label{subst}
G_{\partial_4A_4P}(\vec{p}= 0, t) {}_{\stackrel{\longrightarrow}{t
\rightarrow \infty}} m_{\pi} G_{A_4P}(\vec{p}= 0, t).
\end{equation}
Consequently, Eq.~(\ref{ZA}) becomes
\begin{equation}
Z_A = \lim_{m \rightarrow 0, t \rightarrow \infty} \frac{2 m G_{PP}(\vec{p}= 0, t)}
{m_{\pi} G_{A_4P}(\vec{p}= 0, t)}.
\end{equation}

Since the $G_{PP}$ and $G_{A_4P}$ correlators are calculated at
finite $ma$, we shall define the renormalization factor \be
\label{ZA1} \tilde{Z}_A(ma) = \lim_{t \rightarrow \infty} \frac{2
m G_{PP}(\vec{p}= 0, t)} {m_{\pi} G_{A_4P}(\vec{p}= 0, t)}, \ee
where the massless limit would give $Z_A$. We plot the results of
$\tilde{Z}_A(ma)$ from Eq.~(\ref{ZA1}) in Fig.~\ref{Zaawi}.  In
view of the fact that there is no $O(a)$ error with the overlap
fermion, $\tilde{Z}_A(ma)$ could have terms like $m \Lambda_{QCD}
a^2$ and $m^2 a^2$, but could also have terms like
$m/\Lambda_{QCD}$ and $(m/\Lambda_{QCD})^2$. Thus, we parameterize
it with the form which is linear and quadratic in
$m$~\cite{kent2,liu02} \be  \label{ZAma} \tilde{Z}_A(ma) = Z_A (1
+ b_A m + c_A m^2). \ee As noticed before~\cite{kent2}, it is
conspicuously flat as a function of $ma$ suggesting that the
$O(a^2)$ error from the action and the axial-vector operator is
small. From the fitting to the form in Eq.~(\ref{ZAma}) for the
range of $ma$ from 0.014 to 0.6, we find that $Z_A = 1.849(4), b_A
= -0.347(16)$ (in units of $\Lambda_{QCD}a^2$ with  $\Lambda_{QCD}
= 0.238$ GeV), and $c_A = 0.317(21)$ (in units of $a^2$) with
$\chi^2/dof = 0.54$.

%we find that $Z_A = 1.853(5), b_A = -0.174/1.85/0.25(31/1.85/.25)$ GeV,
%$c_A = 0.330(24)$ with $\chi^2/dof = 0.728$

We observe that $Z_A$ is determined to the precision of 0.2\% in statistical error. It is thus
more desirable~\cite{Tblum1} to use $Z_A$ from the Ward identity and $\Gamma_A$ to determine
all the other renormalization constants.

From the renormalization condition Eq.~(\ref{Z_A}), we finally
obtain $Z_{\psi}^{\rm SI}$
\be Z_{\psi}^{\rm SI} = Z_A
\Gamma_A^{\rm SI}, \ee and the other SI renormalization constants
from Eq.~(\ref{3rd}) \be  \label{SI_ren} Z_O^{\rm SI} = Z_A
\frac{\Gamma_A^{\rm SI}}{\Gamma_O^{\rm SI}}. \ee

Since we normalize the scale invariant vertex functions to the RI scheme at
$(pa)^2 = 4.1$, the renormalization constant determined in Eq.~(\ref{SI_ren}) is just
$Z_O^{RI}(\mu = 2~\rm{GeV})$.

\subsection{Matching to $ {\overline{\rm MS}} $ scheme }

In order to confront experiments, one frequently likes to quote the final
results in the $\overline{\rm MS}$ scheme at certain scale. For light hadrons,
the popular scale is 2 GeV. To obtain the renormalization constants in
the $\overline{\rm MS}$ scheme at 2 GeV, one can use the perturbatively
computed coefficient functions in Eq.~(\ref{c(x)}) in the $\overline{\rm MS}$ scheme
to evolve the scale invariant renormalization constant to the targeted scale, i.e.
\be   \label{SI-MS}
Z_O^{\overline{\rm MS}}(\mu) = C_O^{\overline{\rm MS}}(\mu) Z_O^{\rm SI}.
\ee

Alternatively, one can avoid the step of going to the scale invariant quantity and,
instead, match directly from the RI-scheme to the $\overline{\rm MS}$ scheme at the same scale.
The perturbative expansion of the ratio $Z^{\overline{\rm MS}}/Z^{RI}$ to two-loop order are given
by the finite coefficients in the perturbative expansion of $Z^{RI}$~\cite{Franco:1998bm}
\begin{equation}   \label{eq:R}
R = \frac{Z^{\overline{\rm MS}}}{Z^{RI}} = 1 + \frac{\alpha_s}{4 \pi} (Z^{RI})_0^{(1)}
+ (\frac{\alpha_s}{4 \pi})^2 (Z^{RI})_0^{(2)}  + ....
\end{equation}
The numerical values of the
matching coefficients, $Z^{(1)}_0$ , $Z^{(2)}_0$ and $Z^{(3)}_0$ in Eq.~(\ref{eq:R})
used for $Z_{\psi}$ , $Z_m$ and $Z_T$ have been calculated in Ref.~\cite{Chetyrkin:20004l} and Ref.~\cite{Gracey0},
and  are collected in Table~\ref{tab:match}.

%Jianbo: please verify that $Z_T^{\overline{\rm MS}}$ are indeed obtained from what I describe above.

%### Modify the table to give $Z_A$ from the Ward identity, $Z_{\psi}$ and all the
%other Z's in terms of $Z_{\psi}$ in the RI scheme, the SI scheme and $\overline{\rm MS}$ scheme.
%Do not use ratios (i.e. $Z_S/Z_{\psi}$)
%Jianbo: please add SI results. Please use $Z_A$ from Shao-Jing which is 1.849(4).
%Jianbo: please verify that $Z_T^{\overline{\rm MS}}$ are indeed obtained from what I describe above.
%### Modify the table to give $Z_A$ from the Ward identity, $Z_{\psi}$ and all the
%other Z's in terms of $Z_{\psi}$ in the RI scheme, the SI scheme and $\overline{\rm MS}$ scheme.
%Do not use ratios (i.e. $Z_S/Z_{\psi}$)

%Jianbo: please add SI results. Please use $Z_A$ from Shao-Jing which is 1.849(4).

\begin{table}[!ht]
\begin{center}
\caption{Quenched RI to ${\overline{\rm MS}}$ matching  coefficients }
\label{tab:match}
\begin{ruledtabular}
\begin{tabular}{ccccc}
 $Z_{(0)}$ & $Z_0^{(1)}$ &  $Z_0^{(2)}$ &  $Z_0^{(3)}$  &  R at 2 GeV \\
\hline
$Z_\psi $ & 0.0000     &    -25.4642     &   -1489.9805 &  0.98706 \\
$Z_m  $ & -5.3333   &      -149.0402   &     -5598.9526 &  0.85127\\
$Z_T $ &  0.0000     &     -46.6654    &     -2067.9753 &  0.97909\\
\end{tabular}
\end{ruledtabular}
\end{center}
\end{table}

%The final results for the renormalization constants are listed in Table~\ref{zresult}.
The results for $Z_A, Z_V, Z_P, Z_S, Z_T$ and $Z_{\psi}$ in the RI and $\overline{\rm MS}$ scheme
at 2 GeV (for $Z_P, Z_S, Z_T$ and $Z_{\psi}$) are listed in Table \ref{zresult}.

\begin{table}[ht]
\caption{\label{zresult} Renormalization constants $Z$ in RI, and ${\overline{\rm MS}}$ schemes at
$\mu = 2$ GeV. The renormalization constant in the SI scheme is normalized to be the
same as that in the RI scheme at 2 GeV. These renormalization constants are obtained from the lattice
with $a = 0.200$ fm.}
\begin{ruledtabular}
\begin{tabular}{ccc}
%\hline
$Z$  & RI at 2 GeV & ${\overline{\rm MS}}$ at 2 GeV       \\
\hline
$Z_A $   & 1.853(9)  & 1.853(9)     \\
$Z_V$    & 1.846(8)  & 1.846(8)      \\
$Z_P $   & 1.571(15)  & 1.845(17)   \\
$Z_S $   & 1.567(13)  & 1.841(15)    \\
$Z_T$    & 1.966(6)   & 1.925(6)  \\
$Z_\psi$ & 2.307(18)   & 2.277(18)   \\
%\hline
\end{tabular}
\end{ruledtabular}
\end{table}

\section{Finite $m$ dependence of the renormalization factors}
\label{sec:MA}

As discussed in Sec.~\ref{sec:overlap}, a non-perturbative renormalization of the heavy-light
axial current via the chiral Ward identity and the unequal mass Gell-Mann-Oakes-Renner
relation is possible with the overlap fermion at finite $ma$~\cite{liu02}. This offers an
opportunity to calculate heavy-light decay constants and transition matrix elements without
being subjected to the uncertainty of the perturbative calculation of the renormalization constants.
In this section, we shall examine the finite $m$ behavior of the renormalization factors.
This is useful for the future study of heavy-light decays and transitions.

Similar to the renormalization factor $\tilde{Z}_A(ma)$ defined in
Eqs.~(\ref{ZA1}) and (\ref{ZAma}) for the axial current, we shall
parameterize the renormalization factor for the other operators by
\begin{equation}  \label{fma}
\tilde{Z}_{O} (ma)  = Z_O(\mu a, g(a))(1+b_O m + c_O m^2),
\end{equation}
where the linear $m$ term includes terms like $m \Lambda_{QCD} a^2$ and $m/\Lambda_{QCD}$
and the quadratic $m$ term includes terms like $m^2 a^2$ and $(m/\Lambda_{QCD})^2$.

\begin{figure}[h]
\includegraphics[height=8.0cm]{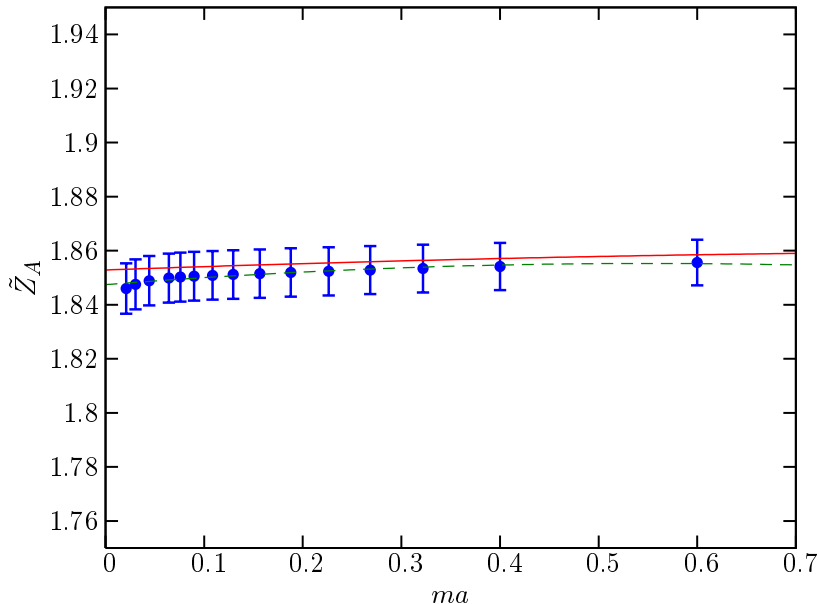}
%\parbox{130mm}{
\caption{Renormalization factor $\tilde{Z_A}(ma)$ against quark mass $ma$. The solid line
is the correlated fit and the dashed line is the uncorrelated fit. }
\label{figvZAm}
\end{figure}

\begin{figure}[h]
\includegraphics[height=8.0cm]{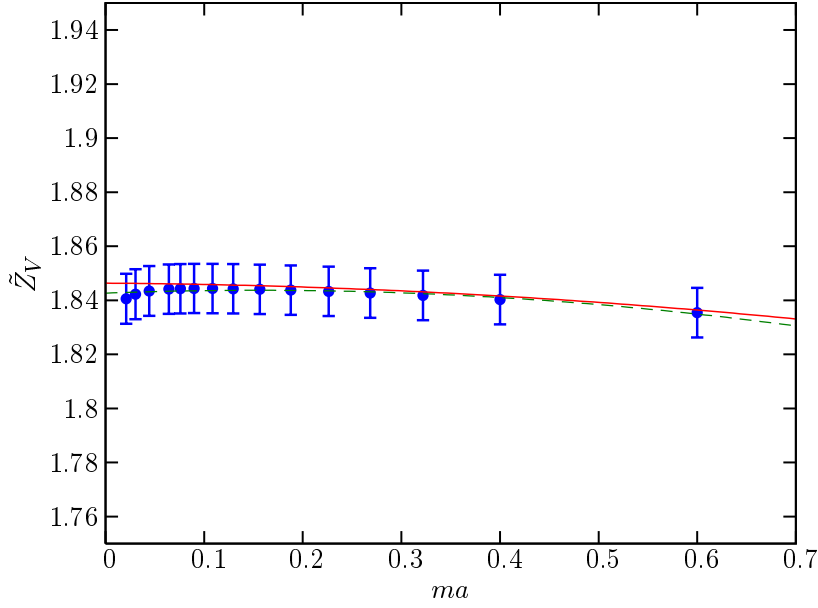}
%\parbox{130mm}{
\caption{The same as in Fig.~\ref{figvZAm} for $\tilde{Z_V}(ma)$. }
\label{figvZVm}
\end{figure}

\begin{figure}[h]
\includegraphics[height=8.0cm]{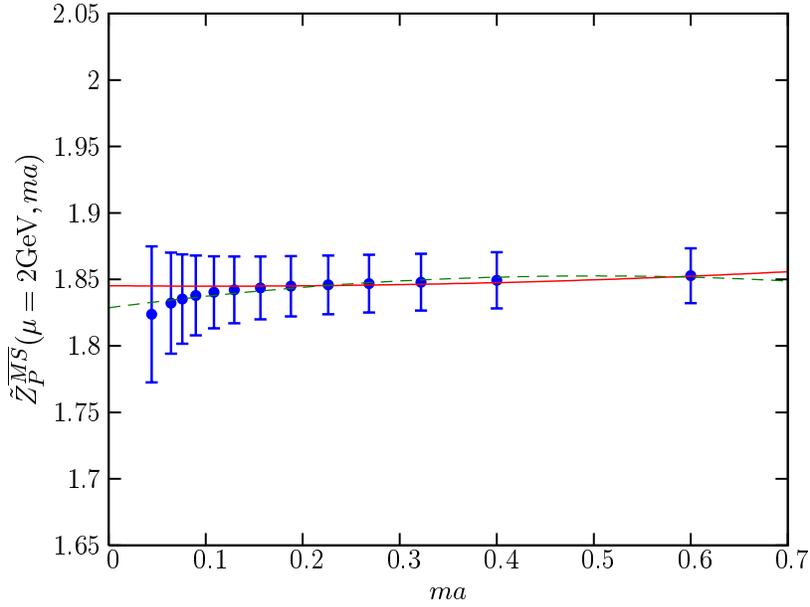}
%\parbox{130mm}{
\caption{The same as in Fig.~\ref{figvZAm} for the renormalization factor
$\tilde{Z_P}^{\overline{\rm MS}}(ma)$  in the $\overline{\rm MS}$ scheme at $\mu =2$ GeV. }
\label{figZPm}
\end{figure}

\begin{figure}[h]
\includegraphics[height=8.0cm]{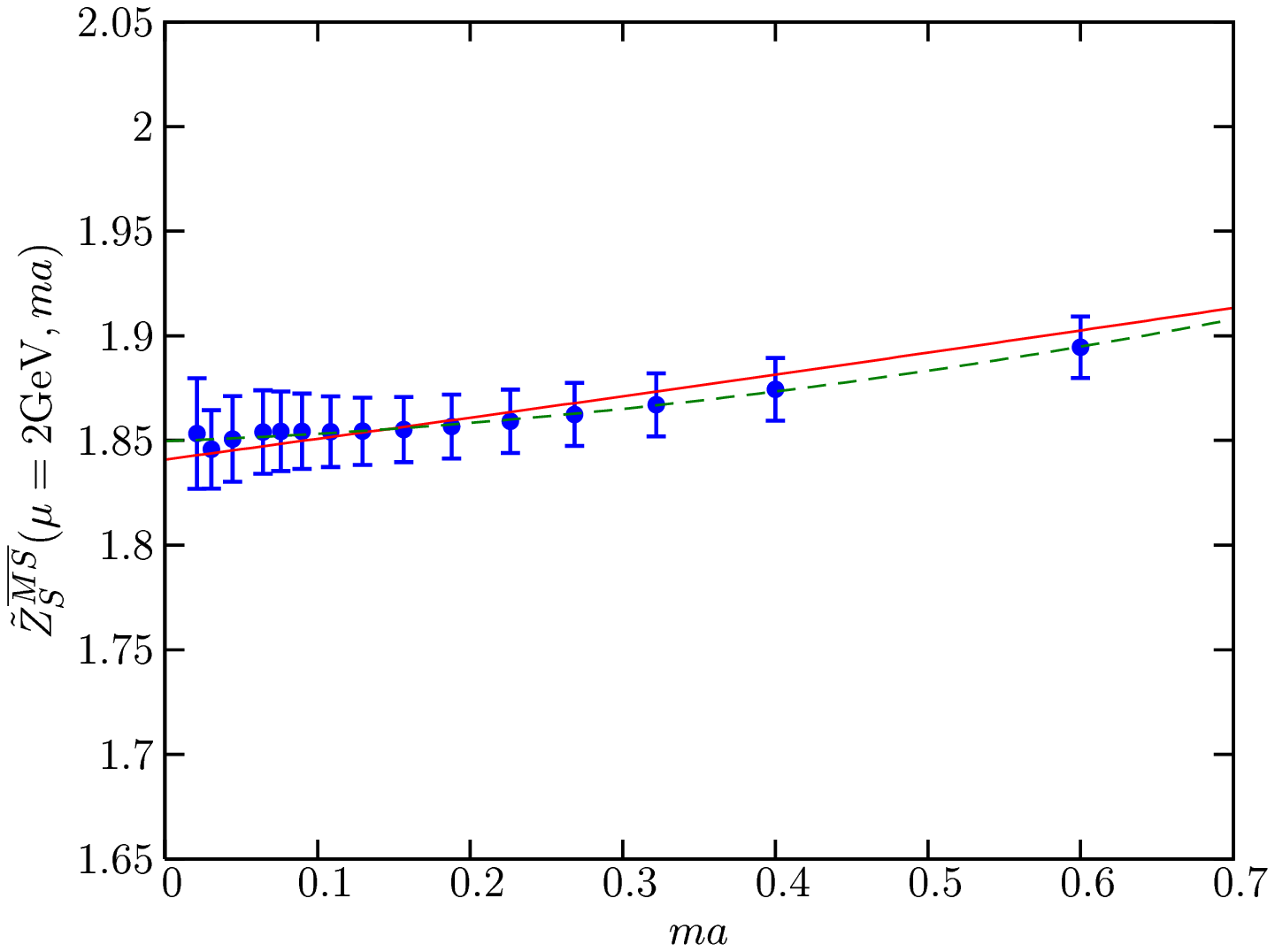}
%\parbox{130mm}{
\caption{The same as in Fig.~\ref{figZPm} for $\tilde{Z_S}^{\overline{\rm MS}}(ma)$. }
\label{figZSm}
\end{figure}

\begin{figure}[h]
\includegraphics[height=8.0cm]{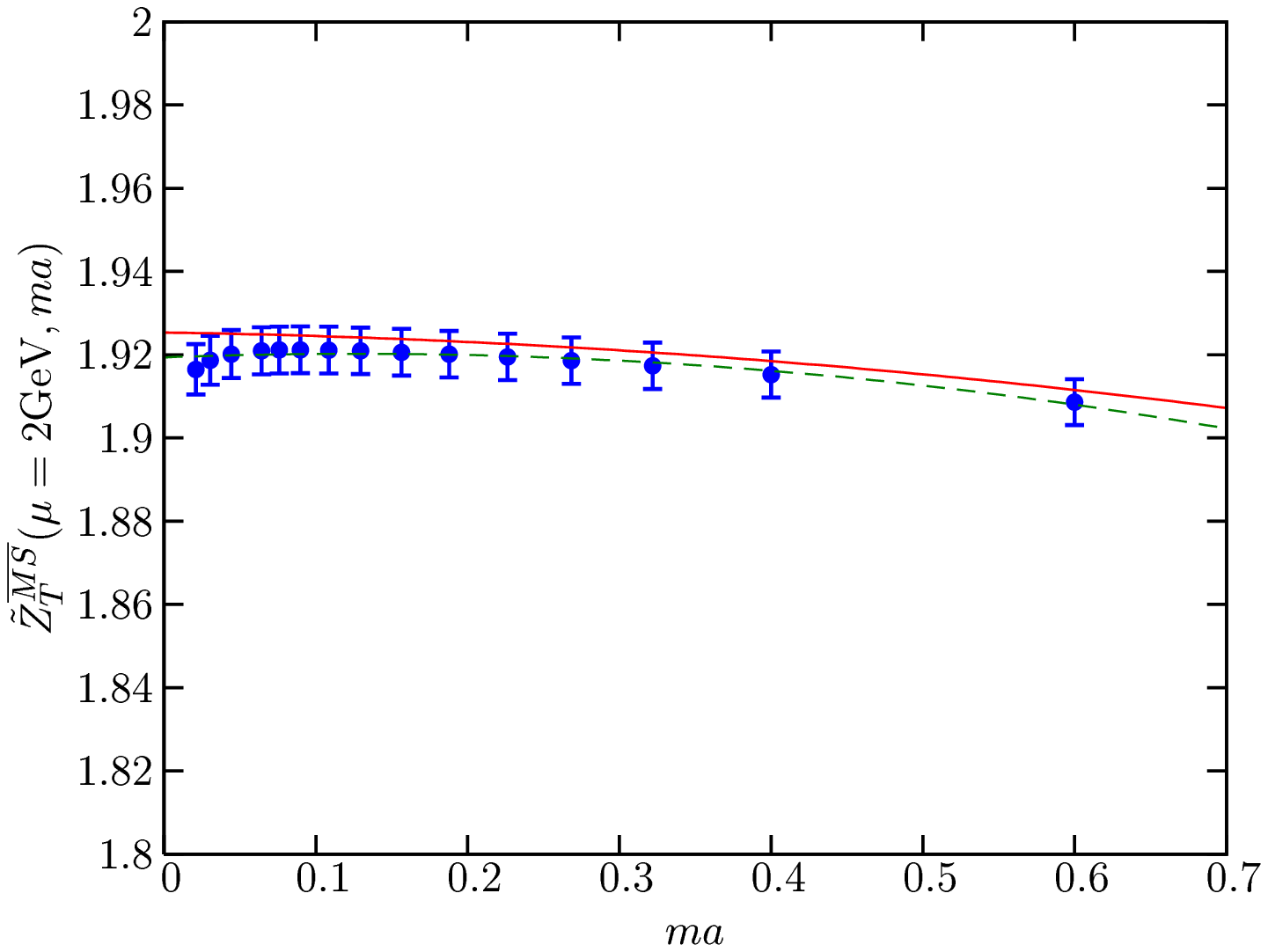}
%\parbox{130mm}{
\caption{The same as in Fig.~\ref{figZPm} for $\tilde{Z_T}^{\overline{\rm MS}}(ma)$. }
\label{figZTm}
\end{figure}

\begin{figure}[h]
\includegraphics[height=8.0cm]{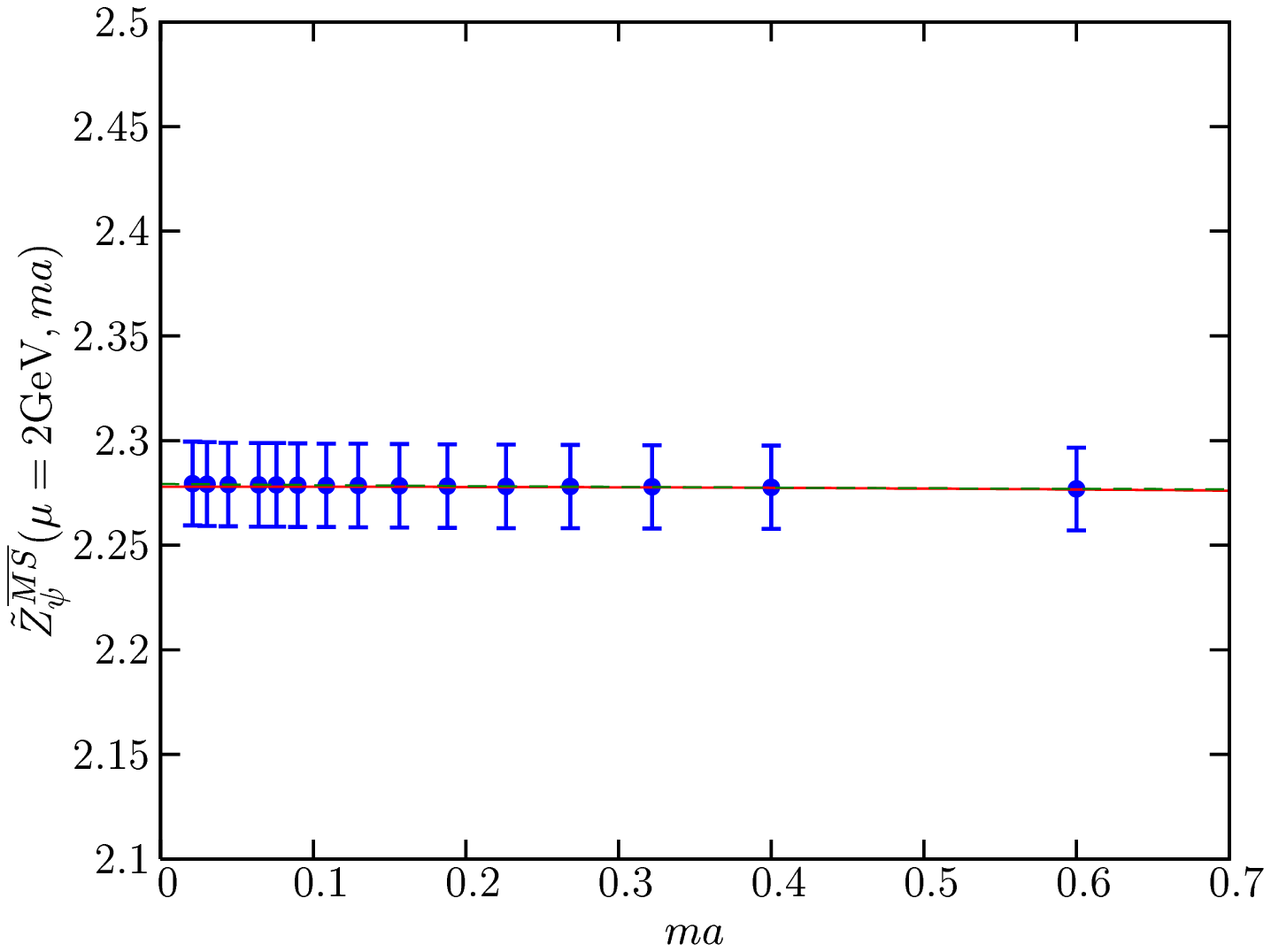}
%\parbox{130mm}{
\caption{The same as in Fig.~\ref{figZPm} for $\tilde{Z_{\psi}}^{\overline{\rm MS}}(ma)$. }
\label{figZfm}
\end{figure}

%$b_O$ has a dimension of $m$ can be written
%as $b_O' \Lambda_{QCD}$ so that $b_O'$ is dimensionless.
%For $Z_A$, the mass dependence is calculated by Axial Ward identity and is plotted
%in Fig.(\ref{Zaawi}).
To obtain the renormalization factors for $ O = V, S, P, T$ and $\psi$, we shall
adopt the same renormalization condition Eq.~(\ref{eq:ren_con}) for finite $m$.
We shall use the quark propagator to obtain the finite $m$ dependence for $\tilde{Z}_{\psi}(ma)$
from Eq.~(\ref{wf_ren}), except with $Z_{\psi} = \tilde{Z}_{\psi}(ma=0)$ normalized
from $Z_A$ via the renormalization condition, Eq.~(\ref{Z_A}). We then follow
the above procedure in the preceding sections to obtain
$\tilde{Z}_A(ma), \tilde{Z}_V(ma), \tilde{Z}_S(ma),
\tilde{Z}_P(ma)$, and $\tilde{Z}_T(ma)$.
The $\tilde{Z}_{\psi}^{\overline{\rm MS}}(ma, \mu = 2~{\rm GeV}$) for
$\tilde{Z}_{\psi}(ma)$ in the $\overline{\rm MS}$ scheme at $\mu = 2$ GeV is
plotted in Fig.~\ref{figZfm}. The corresponding results on
$\tilde{Z}_A(ma),\tilde{Z}_V(ma), \tilde{Z}_S^{\overline{\rm MS}}(ma, \mu = 2~{\rm GeV}),
\tilde{Z}_P^{\overline{\rm MS}}(ma, \mu = 2~{\rm GeV})$ and
$\tilde{Z}_T^{\overline{\rm MS}}(ma, \mu = 2~{\rm GeV})$ are plotted in
Fig.~\ref{figvZAm}, \ref{figvZVm}, \ref{figZSm}, \ref{figZPm} and \ref{figZTm}
respectively.

We observe that these curves are all rather flat, with less than
3\% deviation for $ma$ as large as 0.6. We make a correlated fit
of the finite $m$ behavior with the form in Eq.~(\ref{fma}). The
coefficients $b_O$ and $c_O$ are listed in Table \ref{fit-ma}. The
fitted curves are drawn in Figs.~\ref{figvZAm}, \ref{figvZVm},
\ref{figZSm}, \ref{figZPm} and \ref{figZTm} as solid lines. For
comparison, we also plot the uncorrelated fits as dashed lines
which have much smaller $\chi^2/dof$ than those of the correlated
fits. The flatness in $ma$ suggests that the $O((ma)^2)$
dependence is weak. This is consistent with the findings for the
$\pi$ and $\rho$ masses~\cite{kent1} and their dispersion
relations~\cite{liu02}. The observed $O(ma^2)$ and $O(m^2a^2)$
dependence are much weaker than those found with the chirally
improved Dirac operator~\cite{ggh04}. We don't know the finite $m$
behavior of the domain-wall fermion from the study of the quark
bilinear operators~\cite{Tblum1}. The study of $\tilde{Z}_A (ma)$
and $\tilde{Z}_V (ma)$ from the nucleon matrix elements found much
stronger $O(ma^2)$ and $O(m^2a^2)$ dependence than observed here.

We should point out that in Sec.\ref{sec:ZA}, $\tilde{Z}_A(ma)$
was defined in Eqs.~(\ref{ZA1}) and (\ref{ZAma}) with the implicit
assumption that $\tilde{Z}_P(ma)$ is not very different from
$\tilde{Z}_S(ma)$ for the available range of $ma$. Even though
this does not in principle affect the extraction of the
renormalization constant $Z_A$ at the massless limit of
$\tilde{Z}_A(ma)$, where $Z_P$ is equal to $Z_S$, the
extrapolation could have a large uncertainty if $\tilde{Z}_P(ma)$
is very different from $\tilde{Z}_S(ma)$ for the range of $ma$
which is not close to the chiral limit. Thus, it is gratifying to
see in  Figs. \ref{figZPm} and \ref{figZSm} that the $ma$
behaviors of $\tilde{Z}_P(ma)$ and $\tilde{Z}_S(ma)$ (in the
$\overline{\rm MS}$ scheme at 2 GeV in this case) are very
similar. They differ less than 3\% for $ma \le 0.6$. The upshot is
that we should take the $ma$ behavior in $\tilde{Z}_A(ma)$ in
Fig.~\ref{figvZAm} as the correct one, instead of that in
Fig.~\ref{Zaawi} as defined in Eq.~(\ref{ZA1}). Any attempt to
correct for the negligence of the $ma$ dependence in
$\tilde{Z}_P(ma)$ and $\tilde{Z}_S(ma)$ in Eq.~(\ref{ZA1}) and
compare with that in $\tilde{Z}_A(ma)$ in Fig.~\ref{figvZAm} is
expected to reflect different finite $ma$ corrections under
different renormalization conditions.

\begin{table}[t]
\caption{\label{fit-ma}Mass dependence of the renormalization factors as defined in Eq. (\ref{fma}).
$b_O$ is in units of $\Lambda_{QCD}a^2$ with  $\Lambda_{QCD} = 0.238$ GeV. $c_O$ is in units of $a^2$.}
\begin{ruledtabular}
\begin{tabular}{ccccc}
    & $Z_O^{\overline{\rm MS}}$ (2 GeV) & $b_O $& $c_O$  & $\chi^2/dof$\\
\hline
A   &     1.853(9)  &  0.029(16)       & -0.003(2) &  1.3   \\
V   &     1.846(8)  & -0.0055(33)   & -0.013(7)   & 2.2\\
P   &     1.845(20)  & -0.015(12)  &    0.017(4)  & 0.3   \\
S   &     1.841(15) & 0.22(19)    & 0.004(2)   & 0.7\\
T   &     1.925(6)  & -0.012(8)    & -0.015(3)  & 1.6\\
$\psi$ &  2.277(18) & -0.001(2)    &  0.002(1)  & 0.8      \\
\hline
\end{tabular}
\end{ruledtabular}
\end{table}

%NEW Figure for comparing two $\tilde{Z}_A(ma)$.

\section{summary and outlook}
\label{sec:conclusions}

 In this work, we performed a non-perturbative renormalization calculation of the
composite quark bilinear operators with the overlap fermion in the
regularization-independent scheme from the quark vertex function
with high virtuality. The renormalization group running of the
renormalization constants were calculated to obtain the scale
invariant (SI) renormalization constants and also matched to the
$\overline{\rm MS}$ scheme at $\mu = 2$ GeV. The scale invariant
$Z_A$ from the Ward identity and the axial vertex function were
used to eliminate the wavefunction renormalization $Z_{\psi}$ and
determine the renormalization constants from other vertex
functions. Since $Z_A$, as obtained from the Ward identity, has a
very small error ($\sim 0.2$\%), it is more desirable to use it to
determine $Z_{\psi}$ instead of using the quark propagator to
determine it. The latter can introduce an error as large as $\sim$
10 -- 20\%~\cite{Tblum1}.
%In the current case,
%the difference between the $Z_{\psi}$ calculated from the quark propagator and
%that calculated using $Z_A$ is about 4\%.

   After subtracting the quenched chiral log divergences in the vertex functions
$\Gamma_P$ and $\Gamma_S$ due to the presence of the pseudoscalar
meson, the expected relation $Z_S = Z_P$ due to
chiral symmetry holds to high precision ($\sim  1\%$) for a large range of
$(pa)^2$ with $(pa)^2 > 3$. The same is true for the relation $Z_A = Z_V$. The
resultant check on the chiral symmetry relations are
comparable to those of the domain-wall fermion~\cite{Tblum1} and
somewhat better than the chirally improved fermion~\cite{ggh04} where it is found
that $Z_A/Z_V \sim 1.03$ and $Z_P/Z_S \sim 0.95$ for their smallest
lattice spacing at $a = 0.078$ fm.

  We studied the finite $m$ behavior in the renormalization factors of these
composite operators.
This is useful if one wants to use the same overlap Dirac operator for both the
light and heavy quarks. With present-day computers, it is not practical
to reach a lattice spacing such that $ma \ll 1$ for the charm and bottom
quarks. As such, one would like to have a Dirac operator which has chiral
symmetry and, at the same time, has small $O(ma^2)$ and $O(m^2a^2)$ errors.
It is suggested~\cite{liu02} that the overlap fermion with its effective
quark propagator having the continuum form might be suitable for this
purpose. Indeed the $O(ma^2)$ and $O(m^2a^2)$ errors in the dispersion
relation are shown to be small~\cite{liu02}.
In this study, we find that the finite $m$ dependence is quite gentle in
$\tilde{Z}_{\psi}(ma), \tilde{Z}_A(ma), \tilde{Z}_V(ma), \tilde{Z}_P(ma),
\tilde{Z}_S(ma)$, and $\tilde{Z}_T(ma)$. For $ma$ as
large as 0.6, the deviations are generally less than 3.5\% in these
renormalization factors we studied.

   Since the lattice community is geared to carrying out large scale dynamical fermion calculations
with chiral fermions, it is worth studying the $(pa)^2$ errors in the scale invariant vertices
to gauge the $O(a^2)$ errors and also the mass dependence of the renormalization factors with the
overlap fermion.

\section{Acknowledgment}
Support for this research from the Australian Research Council
and DOE Grants DE-FG05-84ER40154 and DE-FG02-02ER45967 are gratefully acknowledged.

\end{document}